\documentclass[aps,prd,twocolumn,preprintnumbers,amsmath,amssymb,superscriptaddress,groupedaddress,floatfix,nofootinbib]{revtex4}

\usepackage[normalem]{ulem}
\usepackage[utf8]{inputenc}
\usepackage{epsfig,psfrag,graphics,verbatim}
\usepackage{graphicx}
\usepackage[T1]{fontenc}
\usepackage{subfigure}
\usepackage{dcolumn}
\usepackage{bm}
\usepackage{xcolor}
\usepackage{float}
\usepackage{adjustbox}
\usepackage{amsmath}
\usepackage{subcaption}
\usepackage{makecell}
\usepackage{eucal}
\usepackage{varioref}
\usepackage{hyperref}
\usepackage{cleveref}
\usepackage{soul}

\date{March 2025} 

\begin{document}
\title{Testing \( f(R) \)-gravity models with DESI
DR2 2025-BAO and other cosmological data}

\author{Francisco Plaza}
\email{Fran22@fcaglp.unlp.edu.ar}
\affiliation{%
 Facultad de Ciencias Astronómicas y Geofísicas, Universidad Nacional de La Plata, Observatorio Astronómico, Paseo del Bosque,\\
B1900FWA La Plata, Argentina \\
}%
\affiliation{Consejo Nacional de Investigaciones Científicas y Técnicas (CONICET), Godoy Cruz 2290, 1425, Ciudad Autónoma de Buenos Aires, Argentina}
\author{Lucila Kraiselburd}
\affiliation{%
 Facultad de Ciencias Astronómicas y Geofísicas, Universidad Nacional de La Plata, Observatorio Astronómico, Paseo del Bosque,\\
B1900FWA La Plata, Argentina \\
}%
\affiliation{Consejo Nacional de Investigaciones Científicas y Técnicas (CONICET), Godoy Cruz 2290, 1425, Ciudad Autónoma de Buenos Aires, Argentina}

\begin{abstract}
Motivated by the new BAO data and the significant results recently published by the DESI DR2 Collaboration, this study presents a Markov Chain Monte Carlo (MCMC) analysis of all currently viable \( f(R) \) models using this new dataset, and compares its constraining power with that of previous BAO compilations. A corresponding Bayesian model comparison is then carried out. The results reveal, for the first time, very strong statistical evidence in favor of \( f(R) \) models over the standard \(\Lambda\)CDM scenario. The analysis also incorporates data from cosmic chronometers and the latest Pantheon++SH0ES supernovae compilation.

\end{abstract}
\maketitle
\section{Introduction}
 
It has been known since 1998 that the universe is currently undergoing an accelerated expansion \citep{rii1998AJ....116.1009R,Perlmutter1999ApJ...517..565P}. This discovery led to the introduction of the cosmological constant in the standard model of cosmology, through its inclusion in the Einstein-Hilbert action to explain such accelerated expansion of the universe:
\begin{equation}
S=\frac{1}{2\kappa^2}\int d^4x \sqrt{-g}(R-2\Lambda) + S_m(g_{\mu\nu},\psi)
\end{equation}
where \(\kappa^2 = \frac{8\pi G}{c^4}\)

From this action, one obtains Einstein's field equations with the cosmological constant:
\begin{equation}
R_{\mu\nu}-\frac{1}{2}Rg_{\mu\nu}+\Lambda g_{\mu\nu} = \kappa^2 T_{\mu\nu}
\end{equation}

These field equations have successfully described this latest evolutionary stage of the universe. However, some concerns like the Hubble tension between Riess et al. \citep{3riess2024IAUS..376...15R} and Planck Collaboration \cite{Planckcosmo2018} estimates, and the lack of understanding of the physical cause behind the accelerated expansion continue to motivate further studies on this phenomenon.
This last issue arises from the ambiguity in the context of general relativity regarding whether the cosmological constant belongs to the right-hand side of the field equations, implying new forms of matter to be included in $T_{\mu\nu}$, or whether it belongs to the left-hand side, representing previously unknown aspects of spacetime.

From this interpretative problem, several proposals have emerged. Notably, if exotic matter were responsible for this phenomenon (for example, quintessence or k-essence fields \cite{joyce2016ARNPS..66...95J,suji2013CQGra..30u4003T,2006copelandIJMPD..15.1753C,2005quintaPhRvL..95n1301C}), it would constitute approximately 70\% of the universe. Other proposals consist in assuming that the vacuum energy of quantum fields in the standard model of particle physics is the cause of the accelerated expansion and include it in the energy-momentum tensor; however, they provide the most inaccurate prediction in history, deviating by several orders of magnitude \citep{1989weinbergRvMP...61....1W, Bahcall1999, 1995ostrikerNatur.377..600O}.

On the other hand, $f(R)$ theories constitute a promising alternative due to their ontological simplicity, proposing that the expansion is driven by the dynamics of spacetime itself. They represent previously unknown aspects of spacetime dynamics and curvature that become relevant at very large scales, without resorting to vacuum energy or unknown exotic species \citep{Starobinsky1980}. In this way, these theories provide a well-defined and plausible explanation of the physics underlying the epochs of accelerated expansion of the universe—something that remains unclear with the mere inclusion of the cosmological constant and its interpretation within the framework of standard cosmology. These unknown aspects of the gravitational interaction could originate from quantum phenomena intrinsic to gravity itself, or from quantum corrections required for it. Such effects are contemplated as part of an effective theory in this context, representing the low-energy limit of a more fundamental quantum theory of gravitation, or a unified theory of all fundamental interactions \citep{Starobinsky1980,staroycuerdas10.1093/ptep/ptw161,grandi2024PhRvD.110h4043C}. 
Additionally, they are of significant theoretical interest, as they offer a framework in which both inflation and late-time acceleration may be unified within a single underlying mechanism \citep{Nojiri:2008nt,Starobinsky1980,Artymowski:2014gea,hawkingPhysRevD.63.083504,defeliytsuji2010LRR....13....3D}.
This is not achievable within the standard cosmological model, in which the use of a cosmological constant is not capable of explaining both stages of accelerated expansion and requires invoking exotic entities—such as the inflaton field—to address early-universe problems, such as the flatness problem and the horizon problem.

This kind of models arise from a generalization of the action to one that includes a function $f(R)$ instead of just the Ricci scalar:
\begin{equation}
S=\frac{1}{2\kappa^2}\int d^4x \sqrt{-g}f(R) + S_m(g_{\mu\nu},\psi)
\end{equation}

The metric formalism was employed, and the corresponding modified field equations were obtained:
\begin{equation}
f_{R}(R)R_{\mu\nu}-\frac{1}{2}f(R)g_{\mu\nu}-(\nabla_\mu\nabla_\nu-g_{\mu\nu}\Box)f_R(R) = \kappa^2 T_{\mu\nu}
\label{eccampo}
\end{equation}
where ${\it f_R}=\frac{d {\it f}}{d R}$, $\Box$ is the d'Alembertian operator and $\nabla_{\mu}$ is the covariant derivative.
 
In this context, all terms depend on the Ricci scalar, and the accelerated expansion of the universe attributed exclusively to the geometric aspects of spacetime. Even when assuming a vacuum state and $R=constant=R_0$, the dynamics correspond to a universe undergoing accelerated expansion \citep{2023danielaFdeTtgif.book.....A}, characterized by an effective cosmological constant given by:

\begin{equation}
\Lambda_{ef} = \frac{f(R_0)}{ (f_R(R_0))^2}
\end{equation}

This interpretation of $f(R)$ theories remains valid as long as there is no evidence for the hypothetical scalaron, a particle associated with the scalar field that appears when transitioning from the Jordan frame to the Einstein frame \citep{1962dickePhRv..125.2163D,capozziello1997CQGra..14.3243C,capozziello2010PhLB..689..117C}. However, even if the scalaron truly exists, it represents an exotic yet more natural species that arises inherently from the gravitational theory itself and, even some authors dare to suggest the $f(R)$ theories as dark matter candidates as well \citep{Nojiri:2008nt,Park:2023aht}. However, in the present work, dark matter is included as an additional species in the energy-momentum tensor, and it is not modeled as an effect arising from modified gravity. In contrast, dark energy is entirely replaced by these spacetime effects within the framework of \( f(R) \) gravity. If dark matter were not included in the energy-momentum tensor, the model would fail to correctly describe the expansion rate of the universe.

The purpose of this work is to perform statistical analyses to establish bounds on the parameters of $f(R)$ models (Hu-Sawicki, Starobinsky, Exponential and Hyperbolic Tangent) using the most recent data from SnIa (with the inclusion of calibrated-Cepheid supernovae), cosmic chronometers and baryon acoustic oscillations. 
To date, numerous analyses of this kind have been reported \cite{2016MNRAS.459.3880H,Nunes_2017,Odintsov2017,2016PhRvD..93h4016D,M2022PhRvD.105j3526L,ODINTSOV2021115377,furrugia2021PhRvD.104l3503F,2024MNRAS.527.7626R,2024A&A...686A.193K,2025arXiv250205259L}, 
revealing that the earliest fits indicated that the distortion parameter \( b \), which, as we will see in the theoretical formalism of \( f(R) \) gravity, quantifies the deviation of the theory from General Relativity, was often found to be very small. This suggested minimal differences between the theories, as can be seen in \cite{ODINTSOV2021115377,M2022PhRvD.105j3526L,furrugia2021PhRvD.104l3503F}. It was only in 2024 that the inclusion of the Pantheon Plus compilation of Type Ia supernovae in the fits drastically altered the results, yielding higher values for \( b \) and revealing inconsistencies at the 1- to 2-sigma level between the \( f(R) \) model and \(\Lambda\)CDM \citep{plaza2025arXiv250316132P,ravi2024MNRAS.527.7626R,PhysRevD.109.123514}. However, despite these discrepancies, the Bayesian analysis still indicated a statistical preference for the standard model \citep{plaza2025arXiv250316132P,ravi2024MNRAS.527.7626R} or a very slight preference for certain $f(R)$ models.

The recent article from the DESI collaboration \citep{DR22025arXiv250314738D} presents the latest compilation of Baryon Acoustic Oscillation (BAO) measurements and performs a Bayesian comparison between the standard model and simple alternative models. More specifically, with $w_0w_a$CDM models that allow for a dynamical dark energy component—that is, they permit the variation of the equation of state of the presumed dark energy. That study finds statistical preference for such alternative models over $\Lambda$CDM.

This result attracted considerable attention, given that previous evidence against the standard model had never reached statistical significance. This finding motivates the present work, in which we similarly conduct Bayesian comparisons for cosmological models based on $f(R)$ gravity, in contrast with the standard model. It is important to emphasize that the result reported in \citep{DR22025arXiv250314738D} does not imply that dark energy models are necessarily correct, but rather that any model featuring a time-dependent equation of state will be favored by Bayesian criteria. In this regard, models based on $f(R)$ theories also exhibit a time-varying effective equation of state.

However, the interpretations of these models differ dramatically: while the $w_0w_a$CDM model attributes the accelerated expansion to an exotic matter component, $f(R)$ theories—under our interpretation—attribute it to modifications in the gravitational sector, which we consider to be a conceptually simpler approach. Based on these considerations, this work presents, for the first time, an analysis of this kind for models based on $f(R)$ theories of gravity, incorporating data from the DESI DR2 release.

From this, a very strong statistical preference for the $f(R)$ models over the standard model emerges, as will be discussed in the final section.

The article is organized as follows: in Section \ref{Sec:fr} we write the Friedmann equations to be solved to obtain the dynamics of $f(R)$. There is also a summary of the viability conditions that the theory must satisfy and the four models to be analyzed are presented. Next, in Section \ref{Sec:Observational Data} we describe the data chosen to perform the statistical analysis as well as the expressions that relate them to the theory. Our results are presented in Section \ref{Sec:results} and in the following Section, \ref{Sec:discussion}, a comparison is made with previously published estimates. Finally the conclusions are in Section \ref{Sec:conclusions}. The different methods used to solve the dynamic equations are found in Appendix \ref{Sec:Methods}.

\section{Cosmological models in $f(R) $ }
\label{Sec:fr}
The field equations governing gravity in $f(R)$ theories \ref{eccampo}, also admit the FLRW cosmological metric as a solution in terms of the scale factor of the universe $a(t)$:
\begin{equation}
ds^2 = -c^2dt^2 + a^2(t)\left(\frac{dr^2}{1-Kr^2} + r^2d\Omega^2\right)
\end{equation}

The modified Friedmann equations are obtained assuming a spatially flat universe and substituting this solution into equations \ref{eccampo}: 

\begin{equation}
\label{eq:friedmod}
-3H^2 = -\frac{1}{f_R}\left(\frac{8\pi G}{c^4} \rho_i + c^2 \frac{R f_R - f}{2} - 3H\dot{R}f_{RR}\right)
\end{equation}
\begin{equation}
\label{eq:friedmod2}
    -2\dot{H} =
    \frac{1}{f_R}\left[\frac{8\pi G}{c^4}\left(\rho_i + \frac{P_i}{c^2}\right)  + f_{RRR}\dot{R}^2 + \left(\ddot{R} - H\dot{R}\right)f_{RR}\right]
\end{equation}
where $H = \frac{\dot{a}}{a}$ and \(i = m, r\); being $m$ the total matter component, including both dark and baryonic matter, and \(r\) referring to radiation. 
These equations are solved numerically.

\subsection{Viable \(f(R)\) theories }

As is well known, \( f(R) \) functions must satisfy a set of strict requirements \citep{cond12010deto.book.....A,cond22010LRR....13....3D} to ensure that the associated theory meets the stability criteria for its solutions, successfully passes solar system tests, and provides a well-founded cosmological description consistent with observational data on the accelerated expansion of the universe, among other considerations. These conditions can be summarized and enumerated as follows:


\begin{enumerate}
    \item \textbf{Positivity of the effective gravitational constant:}  
    The derivative $f_R = \frac{df}{dR}$ must remain positive for $R \geq R_0 > 0$, where $R_0$ represents the Ricci scalar in the current epoch. This guarantees a positive effective gravitational constant ($G_\text{eff} = G/f_R$) and the absence of unphysical local anti-gravity effects:  
    \begin{equation}
        f_R > 0, \quad \text{for } R \geq R_0 > 0.
    \end{equation}
    
    \item \textbf{Stability under perturbations:}  
    To avoid Dolgov-Kawasaki instabilities, the second derivative $f_{RR} = \frac{d^2f}{dR^2}$ must also remain positive:  
    \begin{equation}
        f_{RR} > 0, \quad \text{for } R \geq R_0.
    \end{equation}
    
    \item \textbf{Asymptotic behavior for large $R$:}  
    Along with the previous condition, any viable model should recover the behavior of the \(\Lambda\)CDM model in the high curvature regime, ensuring compatibility with both late-time cosmology and local tests. This requirement essentially mandates that \(f(R)\) asymptotically approaches the Einstein-Hilbert action for large values of \(R\):  
    \begin{equation}
        f(R) \approx R - 2\Lambda, \quad \text{for } R \gg R_0.
    \end{equation}
    
    \item \textbf{Cosmological dynamics during the matter-dominated era:}  
    The model must satisfy:  
    \begin{equation}
        0 < \frac{R f_{RR}}{f_R}(r) < 1, \quad \text{at } r = -\frac{Rf_R }{f} = -2,
    \end{equation}
    which ensures that the modifications to General Relativity lead to a stable, ghost-free theory, and that the model can reproduce a cosmological evolution that mimics the standard matter-dominated era, followed by a smooth transition to an accelerating phase similar to what is observed in our universe.
\end{enumerate}

To simplify the evaluation of these constraints, $f(R)$ models can be rewritten in a way that explicitly highlights deviations from $\Lambda$CDM. This is achieved by expressing them as:
\begin{equation}
    f(R) = R - 2\Lambda y(R, b, \Lambda),
\end{equation}
where $y(R, b, \Lambda)$ quantifies the deviation from the standard model, and $b$ is the parameter controlling this deviation, the so-called {\it distortion parameter}. This formulation allows for a clearer connection between the modified model and the cosmological constant.

The condition for asymptotic behavior can then be re-expressed as:
\begin{equation}
    \lim_{R \to \infty} f(R) = R + A,
\end{equation}
where $A = -2\Lambda$, ensuring alignment with the $\Lambda$CDM model. This equivalence can be further expressed through the relation:
\begin{equation}
    \Omega^{\Lambda CDM}_{\Lambda,0}(H^{\Lambda CDM}_0)^2 = \Omega^{f(R)}_{\Lambda,0}(H^{f(R)}_{0})^2,
\end{equation}
where the superscripts "\(\Lambda {\rm CDM}\)" or "$f(R)$" denote parameters associated with the \(\Lambda{\rm CDM}\) or \(f(R)\) model, respectively, and the subscript 0 indicates the value at the present epoch. Furthermore, this formulation reveals that, at present, matter and radiation densities remain unchanged between $\Lambda$CDM and $f(R)$ models:
\begin{equation}
    \Omega^{\Lambda CDM}_{i,0}(H^{\Lambda CDM}_0)^2 = \Omega^{f(R)}_{i,0}(H^{f(R)}_0)^2 = \frac{8\pi G}{3}\rho_{i,0}, \quad i = (m, r).
\end{equation}
However, deviations from $\Lambda$CDM arise predominantly during late cosmic times, allowing for alternative explanations of the current accelerated expansion without invoking a fixed cosmological constant.

Parameters such  $\Omega^{\Lambda CDM}_{m,0}$, $b$, $H^{\Lambda CDM}_0$ are used for fitting the models to observational data. Once fitted, the results are re-expressed in terms of $\Omega^{f(R)}_{m,0}$, $b$ and $H^{f(R)}_0$ (from now on $\Omega_{m,0}$, $b$ and $H_0$), which are derived using the equivalence relations mentioned earlier. This approach provides a comprehensive framework for assessing deviations from standard cosmology and their observational signatures.

\subsection{The $f(R)$ theories}


In this subsection, we present four of the most popular $f(R)$ gravity theories that satisfy all requirements outlined in the previous subsection and have been tested to date. Initially, each $f(R)$ function is introduced in its original formulation, followed by its reformulation into a more intuitive representation that aligns with the Einstein-Hilbert action. This approach highlights their potential to approximate the $\Lambda$CDM paradigm under appropriate conditions. The $f(R)$ theories studied in this work are:

\subsubsection*{\textbf{i. Hu-Sawicki}}
The Hu-Sawicki theory, introduced in \citep{husawicki2007PhRvD..76f4004H}, presents a functional form for $f(R)$ gravity aimed at mimicking late-time cosmic acceleration:
\begin{equation}
    f_{\text{HS}}(R) = R - \frac{c_1R_{HS}  \left(\frac{R}{R_{HS}}\right)^{n_{\text{HS}}}}{1 + c_2 \left(\frac{R}{R_{HS}}\right)^{n_{\text{HS}}}},
\end{equation}
where $c_1$ and $c_2$ are dimensionless constants, $n_{\text{HS}}$ is a positive integer, and $R_{HS} \approx  \Omega_{m0} H_0^2$. By redefining parameters as $ c_1 R_{HS} / (2c_2) \equiv \Lambda$ and $b \equiv 2 c_2^{1-1/n_{\text{HS}}} / c_1$, the expression simplifies to:
\begin{equation}
    f_{\text{HS}}(R) = R - 2\Lambda \left\{1 - \left[ 1 + \left( \frac{R}{b\Lambda} \right)^{n_{\text{HS}}} \right]^{-1} \right\}.
\end{equation}

Here, $\Lambda$ represents the cosmological constant, while $b$ quantifies deviations from the $\Lambda$CDM paradigm. Viability conditions require $f_R > 0$ and $f_{RR} > 0$ for $R \geq R_0$, implying $b > 0$ when $n_{\text{HS}}$ is odd.

\subsubsection*{\textbf{ii. Starobinsky}}
Starobinsky's approach \cite{Starobinsky1980} modifies $f(R)$ as:
\begin{equation}
    f_{\text{ST}}(R) = R - \lambda R_S \left[ 1 - \left( 1 + \frac{R^2}{R_S^2} \right)^{-n_S} \right],
\end{equation}
where $n_S > 0$, $\lambda > 0$, and $R_S \approx R_0$. Recasting this model in a comparable form, we define $\Lambda = \lambda R_S / 2$ and $b = 2 / \lambda$:
\begin{equation}
    f_{\text{ST}}(R) = R - 2\Lambda \left\{ 1 - \left[ 1 + \left( \frac{R}{b\Lambda} \right)^2 \right]^{-n_S} \right\}.
\end{equation}

For $n_S = 1$, the Starobinsky theory aligns with the Hu-Sawickitheory at $n_{\text{HS}} = 2$. Unlike the Hu-Sawicki theory, the Starobinsky formulation does not necessitate $b > 0$. Nonetheless, we assume $b > 0$ for simplicity in investigating deviations from $\Lambda$CDM.

\subsubsection*{\textbf{iii. Exponential}}
The exponential theory \citep{expPhysRevD.77.046009} introduces a distinct parameterization:
\begin{equation}
    f_\text{E}(R) = R + \alpha \left[ \exp(-\beta R) - 1 \right],
\end{equation}
where $\alpha > 0$ and $\beta > 0$ ensure viability. Redefining parameters as $\Lambda = \alpha/2$ and $b = 2/(\alpha \beta)$, the function takes the form:
\begin{equation}
    f_\text{E}(R) = R - 2\Lambda \left[ 1 - \exp\left(-\frac{R}{b\Lambda}\right) \right].
\end{equation}

In the limit $R \gg b\Lambda$, the theory asymptotically approaches $f(R) \sim R - 2\Lambda$, reproducing $\Lambda$CDM at high curvatures.

\subsubsection*{\textbf{iv. Hyperbolic Tangent}}
Tsujikawa proposed an alternative \citep{Tsujikawa_2008}:
\begin{equation}
    f_\text{T}(R) = R - \xi R_T \tanh\left(\frac{R}{R_T}\right),
\end{equation}
where $\xi > 0$ and $R_T$ are free parameters. This formulation offers unique phenomenological features distinct from other $f(R)$ models.
The Hyperbolic Tangent theory introduces two key parameters, $\xi > 0$ and $R_T > 0$, which govern the model's behavior. By redefining these parameters in terms of $\Lambda = \xi R_T / 2$ and $b = 2 / \xi$, the functional form of $f(R)$ can be expressed as:
\begin{equation}
    f_\text{T}(R) = R - 2\Lambda \tanh\left(\frac{R}{b\Lambda}\right).
\end{equation}

This expression clearly demonstrates that as $b \to 0$ (corresponding to $\xi \to \infty$ and $R_T \to 0$, while ensuring $\xi R_T$ remains finite), the model asymptotically approaches the $\Lambda{\rm CDM}$ model

\section{Observational Data}
\label{Sec:Observational Data}
In this section we summarize the observational data that we use to estimate the parameters of the ${\it f}(R)$ models presented above.

\subsection{Cosmic chronometers}

The cosmic chronometer (CC) method \cite{simon05} is based on determining values of the Hubble parameter $H(z)$ for different redshifts through the study of the differential age evolution of ancient passive-evolving elliptical galaxies, which formed at the same time but are separated by a small redshift interval. Table \ref{tab:CC} presents estimates of  $H(z)$ obtained with this technique that we consider for this work.

\bgroup
\def\arraystretch{1.3}
\begin{table}
\resizebox{0.3\textwidth}{!}{
\begin{tabular}{| c | c | c | c |}
    \hline
    $z$ & $H(z) \ (\mathrm{km}\,\mathrm{s}^{-1}\,\mathrm{Mpc}^{-1}) $ & Reference \\
    \hline
     0.09 & 69  $\pm$  12   & \\
    0.17 & 83  $\pm$  8    &  \\
    0.27 & 77  $\pm$  14   & \\
    0.4  & 95 $\pm$ 17     & \\
    0.9  &117 $\pm$ 23     &~\cite{simon05}  \\
    1.3  &168 $\pm$ 17     & \\
    1.43 &177 $\pm$ 18     & \\
    1.53 &140 $\pm$ 14     & \\ 
    1.75 &202 $\pm$ 40     & \\
    \hline
    0.48 & 97  $\pm$  62   &~\cite{stern10} \\
    0.88 & 90  $\pm$  40   & \\
    \hline  
    0.1791 &  75  $\pm$  4 & \\ 
    0.1993 & 75  $\pm$  5  & \\
    0.3519 & 83  $\pm$  14 & \\
    0.5929 & 104  $\pm$  13 &~\cite{moresco12} \\
    0.6797 & 92  $\pm$  8  & \\
    0.7812 & 105  $\pm$  12 & \\
    0.8754 & 125  $\pm$  17 & \\
    1.037  & 154  $\pm$  20 & \\
    \hline   
    0.07   & 69  $\pm$  19.6   &  \\
    0.12   & 68.6  $\pm$  26.2 &~\cite{zhang14} \\  
    0.2    & 72.9  $\pm$  29.6 & \\
    0.28   & 88.8  $\pm$  36.6 & \\
    \hline 
    1.363  & 160  $\pm$  33.6  &~\cite{moresco15}  \\
    1.965  & 186.5  $\pm$  50.4 & \\ 
    \hline 
    0.3802 & 83  $\pm$  13.5   &  \\
    0.4004 & 77  $\pm$  10.2   & \\
    0.4247 & 87.1  $\pm$  11.2 &~\cite{CC2}\\
    0.4497 & 92.8  $\pm$  12.9 & \\
    0.4783 & 80.9  $\pm$  9    & \\
    \hline
    0.75 & 98.8  $\pm$  33.6    & \citep{Borghi_2022}\\
\hline 
\end{tabular}}
\caption{$H(z)$ values from the CC. Each column shows the redshift of the measurement, the $H(z)$ mean value (and its standard deviation)  and  reference, respectively.}
\label{tab:CC}
\end{table}

\subsection{Supernovae type Ia}

Among the so-called “cataclysmic variables” there are some particularly energetic ones, the type Ia supernovae (SNeIa), which can be observed at large distances of up to z $\sim$ 2.3 so far. This group is chosen as one of the premier distance estimators due to the homogeneity of the spectra of these objects and their light curves, as well as the number of them distributed in all directions.

The distance module $\mu$ can be inferred both from the SNeIa data (being $m_b$  an overall flux normalization and $M_{abs}$ the absolute magnitude of the star)
\begin{equation}
\mu=m_b-M_{abs}
\label{mu_obs}
\end{equation}
and through the following theoretical expression
\begin{equation}
\mu=25+5 \log_{10}(d_L(z)),
\label{distmod}
\end{equation}
where $d_L$ the luminosity distance
\begin{equation}
d_L(z)=(1+z)\int_0^z\frac{c}{H(z')}dz'.
\label{distlum}
\end{equation}
In this article, 
we consider 1590 data from the Pantheon$^+$ compilation (PP)\cite{2022ApJ...938..110B,2022ApJ...938..113S} with $z>0.01$. 
supplemented with 42 SNeIa calibrated by SH0ES Cepheids of the host galaxies \cite{riess2022ApJ...934L...7R}(PPS). 
The addition of the latter means that it is not necessary
to adopt a method to the analysis of absolute magnitude like the ones described in \cite{Conley,Camarena2020,Camarena2023,2024MNRAS.527.7626R,2023PhRvD.107f3523C,PhysRevD.109.123514}. Such that, the $M_{abs}$ can be taken as a free parameter without having to add any further marginalization on it beyond that provided by SH0ES. Although these results 
have recently been questioned by the scientific community\citep{2024arXiv240806153F,peri2024PhRvD.110l3518P}, they have not been discarded and thus can be taken into account.

\subsection{Baryon acoustic oscillations}
\label{bao}

The Baryonic Acoustic Oscillations (BAO) are a phenomenon that occurred before the decoupling of matter and radiation, when photons and baryons were strongly coupled through Thomson scattering, generating perturbations in the density of baryonic matter. These perturbations undergo oscillations that propagate as pressure waves because, on one hand, the presence of gravitational potentials increases the baryonic and dark matter densities, while on the other hand, the collisions between baryonic matter and radiation generate a decrease. When the temperature of the Universe is low enough for the decoupling of light with matter to occur (a stage known as the drag epoch $z_d$), the acoustic oscillations {\it freeze }, leaving a characteristic mark both in the cosmic microwave background and in the large-scale distribution of matter. The sound horizon at the drag epoch $r_d$ is a characteristic scale given by the maximum distance that the acoustic wave could travel in the plasma before decoupling. It turns out that any overdensity of materia in the primordial plasma served as a seed for the birth of future galaxies. Therefore, it is worth noting that this characteristic scale is not only detected in the  microwave background, but is also detected in the distribution of galaxies at the present time. 
In this way, BAOs can be used as a standard ruler for measuring cosmological distances.

We consider two collections of BAO data points obtained using different observational probes from a range of surveys. In the first one (BAO$_1$)(shown in Table ~\ref{tab:BAO_data}), we have collected 20 data from the SDSS \cite{SDSS_DR7,SDSSIIILaforests,SDSSIVLRGanisotropicCF,SDSSIVQSOanisotropicPS,SDSSIVquasars,BOSS,Laforestsquasarscross} , DES \cite{DES_Y1} and the WiggleZ Dark Energy Survey \cite{WiggleZ} surveys ; and in the other, we use the data set from DESI DR2 release \citep{DR22025arXiv250314738D} (BAO$_2$). In this new data version from the DESI collaboration, the accuracy of the measurements has been improved thanks to a larger dataset
of galaxies and quasars compared to the first version
(DR1) \cite{2024arXiv240403002D}. Furthermore, separate measurements of the
transverse comoving distance $D_M (z)$ and the Hubble distance $D_H (z)$ can now be obtained thanks to the improved
signal-to-noise ratio of the new quasar set, which is a major advance, since only the value of the volume-averaged distance $D_V (z)$ was reported in DR1 and DR2. Something to keep in mind is that both compilations should not be used in their entirety in a single statistical analysis because the sky region and redshift range observed by DESI partially overlap with those previously studied whose observations are part of the BAO$_1$ group. Although
the input catalog data used in the DESI and SDSS BAO analyses are different due to details of instrument performance and observing strategy, there are correlations between some of the shared objects
which are larger at low redshifts. Some of the data between both compilations are in agreement but others do not, and the causes of these inconsistencies are still being discussed (although DESI data are supposed to be more accurate thanks to improvements in fitting processes and in raw data processing). Despite the fat that a single data set can be assembled by interleaving some data from each compilation (details in \cite{2024arXiv240403002D,DR22025arXiv250314738D}), we decided to use each data set separately and analyze the estimates obtained for the different $f(R)$ models separately in order to compare them with each other. 

The different BAO$_1$ observables that are listed in Table ~\ref{tab:BAO_data} \footnote{these distance measurements include the $r_d^{fid}$, the sound horizon at  the drag epoch of each fiducial cosmology used in each observation.}  are defined by,
\begin{itemize}
   \item Hubble distance $D_H(z)$,
\begin{equation}
    D_H(z)=\frac{c}{H(z)}
\end{equation}   
    \item Diameter angular distance $D_A(z)$,
\begin{equation}
    D_A(z) = 
    \frac{1}{(1+z)} \int_0^z \frac{c}{H(z')}dz'
\end{equation}
     \item Transverse comoving distance $D_M(z)$
     \begin{equation}
      D_M(z) = (1+z)D_A(z)
    \end{equation}
     \item Volume-averaged distance $D_V(z)$
     \begin{equation}
      D_V(z) = \left[ (1+z)^2 D_A^2(z) \frac{c z}{H(z)}  \right]^{1/3}.
     \end{equation}
\end{itemize}

\bgroup
\def\arraystretch{1.5}
\begin{table}
\resizebox{0.48\textwidth}{!}{
\begin{tabular}{| c | c | c | c |}
\hline 
 $z_{\rm eff}$ &  Value & Observable  & Reference  \\ 
\Xhline{2.5\arrayrulewidth}
$0.15$	& 	$4.473 \pm 0.159$   &  $D_V/r_d$  &	\cite{SDSS_DR7}\\
\hline
$0.44$	& 	$11.548 \pm 0.559$  &  $D_V/r_d$  &	\\
$0.6$	& 	$14.946 \pm 0.680 $  &  $D_V/r_d$  &	\cite{WiggleZ}\\
$0.73$	& 	$16.931 \pm 0.579$   &  $D_V/r_d$  &	\\
\hline
$1.52$	& 	$26.005 \pm 0.995$   &  $D_V/r_d$  &	\cite{SDSSIVquasars}\\
\Xhline{2.5\arrayrulewidth}
$0.81$	& 	$10.75 \pm 0.43$    &  $D_A/r_d$  &~\cite{DES_Y1}\\
\Xhline{2.5\arrayrulewidth}
$0.38$	& 	$10.272 \pm 0.135 \pm 0.074$   &  $D_M/r_d$    & \\
$0.51$	& 	$13.378 \pm 0.156 \pm 0.095$ &  $D_M/r_d$   &~\cite{BOSS}\\
$0.61$	& 	$ 15.449 \pm 0.189 \pm 0.108$    &  $D_M/r_d$   & \\
\hline
$0.698$	& 	$ 17.65 \pm 0.3$    &  $D_M/r_d$   & \cite{SDSSIVLRGanisotropicCF}\\
\hline
$1.48$	& 	$  30.21 \pm 0.79$    &  $D_M/r_d$   & \cite{SDSSIVQSOanisotropicPS} \\
\hline
$2.3$	& 	$37.77 \pm 2.13$   &  $D_M/r_d$  &		\cite{SDSSIIILaforests}\\
\hline
$2.4$	& 	$36.6 \pm 1.2$  &  $D_M/r_d$  &	\cite{Laforestsquasarscross}\\
\Xhline{2.5\arrayrulewidth}
$0.698$	& 	$ 19.77 \pm 0.47$    &  $D_H/r_d$   & \cite{SDSSIVLRGanisotropicCF}\\
\hline
$1.48$	& 	$  13.23 \pm 0.47$    &  $D_H/r_d$   & \cite{SDSSIVQSOanisotropicPS} \\
\hline
$2.3$	& 	$9.07 \pm 0.31 $  &  $D_H/r_d$  &	\cite{SDSSIIILaforests}\\
\hline
$2.4$	& 	$8.94 \pm 0.22$   &  $D_H/r_d$  &	\cite{Laforestsquasarscross}\\
\Xhline{2.5\arrayrulewidth}
$0.38$	& 	$12044.07 \pm 251.226 \pm 133.002 $  &  $H r_d$ [km/s]  & \\
$0.51$	& 	$ 13374.09 \pm 251.226 \pm 147.78$    &  $H r_d$  [km/s]    & ~\cite{BOSS}\\
$0.61$	& 	$ 14378.994 \pm 266.004 \pm 162.558 $    &  $H r_d$ [km/s]   & \\
\hline
\end{tabular}}
\caption{Distance bounds from BAO measurements of different observational probes and surveys. The table include the effective redshift of the measurement, the mean value and standard deviation, the observable and the  reference.}
\label{tab:BAO_data}
\end{table}

\section{Results}
\label{Sec:results}

Here we present the results obtained for each $f(R)$  model from performing statistical analyzes using a Markov chain Monte Carlo method (emcee Python's library)  and the observational data previously described in section \ref{Sec:Observational Data}. The free parameters to be constrained are: the distortion parameter $b$, the total matter density parameter $\Omega_{m0}$, the Hubble parameter $H_0$ and the absolute magnitud $M_{abs}$, the latter in cases where the  SnIa observational data are used. 

The parameter spaces estimated from the different analyses are detailed in Table  \ref{tab:results1} and Figures \ref{fig:staro0}, \ref{fig:staro}, \ref{fig:hu}, \ref{fig:exp}, \ref{fig:tanh} and \ref{fig:all}. The results obtained for the $\Lambda{\rm CDM}$ model are also added in Table \ref{tab:results0} and Fig. \ref{fig:LCDM}. Our results indicate
that the BAO data restrict the allowed parameter spaces more than the PPS and CC data alone. Furthermore, it is observed that  those of BAO coming from
DESI (BAO$_2$) are more restrictive than those of
BAO$_1$ (mostly being from the 
Sloan Digital Sky Survey), but the estimated spaces from both analyzes are
consistent with each other. These behaviors are reflected both in the four $f(R)$ models studied in this article and in the standard cosmological model as well. On the other hand, the results obtained show that the data from Pantheon$^+$ shift and also increase the allowed parameter space of the distortion parameter $b$  with respect to the previous results obtained using the Pantheon data \cite{M2022PhRvD.105j3526L,ODINTSOV2021115377,furrugia2021PhRvD.104l3503F} and that this fact is not due to the inclusion of the SnIa cepheid-calibrated SH0ES data, as can be seen in Fig. \ref{fig:staro0} for the Starobinsky model but that is also reflected in the other models. The inclusion of the PPS compilation not only enhances the constraining power on the cosmological parameters $M_{abs}$ and $H_0$ (and to a lesser extent on $b$ and $\Omega_{m0}$
 ), but also, as already intuited, shifts the one-dimensional posterior distributions toward higher values of $M_{abs}$ and $H_0$, and lower values of $\Omega_{m0}$. The $b$ parameter issue will be again briefly described in the next section when comparing our findings with those of other authors. In addition, as has already been seen in other articles \cite{furrugia2021PhRvD.104l3503F,2024MNRAS.527.7626R} the intervals obtained for this parameter differ between the $f(R)$ models due to the different functional forms that each one has. However, the allowed estimated regions for the rest of the free parameters are consistent between the $f(R)$ models analyzed here.
Finally, it is important to note that although $f(R)$ models do not seem to alleviate the $H_0$ tension, they do not exhibit a greater tension than the standard model in fits such as the one presented in this work, which relies on late-universe data and the model-independent estimate by Riess\cite{riess2022ApJ...934L...7R}. 

\begin{table*}[t]
\renewcommand{\arraystretch}{2.4}
\hspace{-0.40in}
\centering
\begin{tabular}{lcccc}

         
         \hline
        \hline
        Model&Data&$H_0$&$\Omega_{m0}$&$M_{abs}$\\
        \hline

        $\Lambda{\rm CDM}$&{CC+PPS }&$71.604_{-0.936(1.745)}^{+0.822(1.678)}$&$0.324_{-0.017(0.031)}^{+0.014(0.030)}$&$-19.295_{-0.026(0.052)}^{+0.024(0.047)}$\\
        

        &CC+PPS+BAO$_{1}$&$70.657_{-0.685(1.162)}^{+0.466(1.100)}$&$0.319_{-0.010(0.019)}^{+0.009(0.019)}$&$-19.326_{-0.020(0.036)}^{+0.017(0.037)}$ \\
        
        &CC+PPS+BAO$_{2}$&$70.474_{-0.285(0.716)}^{+0.340(0.770)}$&$0.301_{-0.006(0.015)}^{+0.007(0.015)}$&$-19.338_{-0.010(0.025)}^{+0.012(0.027)}$\\
            

         \hline
        
\end{tabular}
\caption{ $\Lambda{\rm CDM}$ results from statistical analysis using different datasets (Cosmic Chronometers (CC), SneIa from Pantheon$^+$+SH0ES (PPS) Collaborations,
and several datasets from  BAOs (BAO$_1$ and BAO$_2$)). For each parameter, we present the mean
value and the $68\%$ ($95\%$) confidence levels or the upper limits obtained.}
\label{tab:results0}
\end{table*}
\begin{figure}
    \centering
    \includegraphics[width=0.95\linewidth]{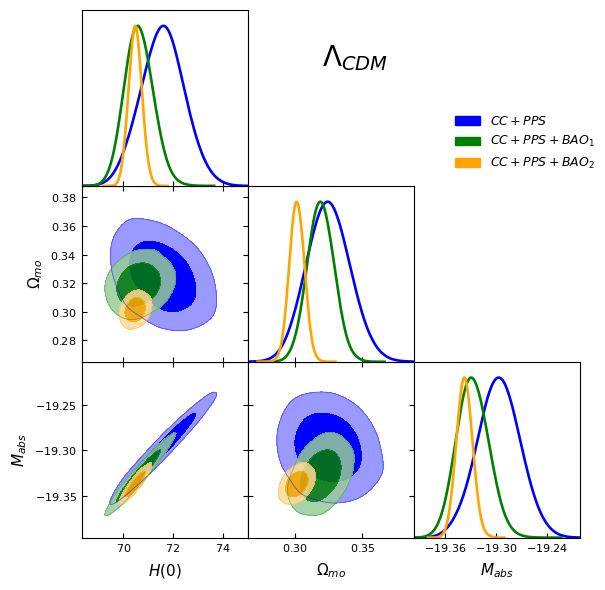}
    \caption{Results for the matter density $\Omega_{m0}$, the Hubble
parameter $H_0$ and the absolute magnitud $M_{abs}$ for the $\Lambda{\rm{CDM}}$ model using combinations of the Cosmic Chronometers, SneIa from Pantheon$^+$+SH0ES  Collaborations, and two different datasets from  BAO (BAO$_1$ and BAO$_2$). The plots show the  $68\%$ and $95\%$ confidence regions.}
    \label{fig:LCDM}
\end{figure}

\begin{table*}[t]
\renewcommand{\arraystretch}{2.4}
\footnotesize
\hspace{-0.40in}
\centering
\begin{tabular}{lccccc} 
         
         \hline
              \hline
          Model&  Data &b&$H_0$&$\Omega_{m0}$&$M_{abs}$\\
            \hline

             Hu-Sawicki& CC+$\rm{PPS}$&$0.660_{-0.297(0.649)}^{+0.315(0.535)}$&$71.430_{-1.109(2.461)}^{+0.645(1.746)}$&$0.261_{-0.011(0.018)}^{+0.038(0.067)}$&$-19.292 _{-0.032(0.068)}^{+0.016(0.044)}$\\
        &CC+$\rm{PPS}$+BAO$_{1}$&$0.061^{+0.155(0.376)}_{-0.048(0.061)}$&$70.369_{-1.102(2.306)}^{+0.502(1.383)}$&$0.318_{-0.011(0.027)}^{+0.010(0.025)}$&$-19.333_{-0.031(0.068)}^{+0.015(0.044)}$\\

        &CC+$\rm{PPS}$+BAO$_{2}$&$0.000066_{-0.000050(0.000066)}^{+0.111(0.317)}$&$70.091_{-0.948(2.036)}^{+0.061(0.424)}$&$0.301_{-0.004(0.012)}^{+0.009(0.018)}$&$-19.349_{-0.026(0.054)}^{+0.003(0.017)}$\\
         \hline
         Starobinsky&CC+PP&$1.847_{-0.865(1.769)}^{+1.159(1.912)}$&$67.374_{-1.343(2.114)}^{+1.758(3.088)}$&$0.302_{-0.038(0.072)}^{+0.024(0.058)}$&$-19.415_{-0.040(0.093)}^{+0.056(0.129)}$\\

        &CC+PPS&$2.498_{-0.970(2.014)}^{+0.799(1.335)}$&$71.464_{-1.018(2.828)}^{+1.064(2.221)}$&$0.267_{-0.024(0.052)}^{+0.033(0.067)}$&$-19.289 _{-0.027(0.065)}^{+0.026(0.062)}$\\


        &CC+$\rm{PPS}$+BAO$_{1}$&$0.531^{+0.520(1.148)}_{-0.354(0.531)}$&$70.169_{-1.058(1.929)}^{+0.545(1.599)}$&$0.318_{-0.010(0.027)}^{+0.012(0.026)}$&$-19.336_{-0.026(0.056)}^{+0.018(0.047)}$\\ 
        &CC+$\rm{PPS}$+BAO$_{2}$&$0.827_{-0.119(0.291)}^{+0.086(0.218)}$&$68.824_{-0.309(0.746)}^{+0.368(0.939)}$&$0.308_{-0.008(0.016)}^{+0.006(0.015)}$&$-19.376_{-0.011(0.025)}^{+0.012(0.029)}$\\
              \hline

         Exponential        
         &CC+$\rm{PPS}$&$2.078^{+0.626(1.175)}_{-0.680(1.645)}$&$71.503_{-1.060(2.913)}^{+1.104(2.913)}$&$0.282_{-0.030(0.066)}^{+0.035(0.065)}$&$-19.288_{-0.027(0.062)}^{+0.026(0.062)}$\\
         
         &CC+$\rm{PPS}$+BAO$_{1}$&$1.298^{+0.437(0.695)}_{-0.513(1.160)}$&$70.073_{-0.744(1.552)}^{+0.765(1.921)}$&$0.320_{-0.010(0.026)}^{+0.012(0.027)}$&$-19.334_{-0.020(0.045)}^{+0.022(0.051)}$\\
         &CC+$\rm{PPS}$+BAO$_{2}$&$1.020_{-0.067(0.183)}^{+0.083(0.193)}$&$69.801_{-0.270(0.654)}^{+0.321(0.754)}$&$0.306_{-0.006(0.015)}^{+0.007(0.017)}$&$-19.351_{-0.010(0.024)}^{+0.011(0.027)}$\\

            \hline
        Hyperbolic tangent&   {CC+$\rm{PPS}$ }&$3.979^{+0.660(1.020)}_{-1.325(2.846)}$&$71.487_{-0.559(1.773)}^{+1.145(2.270)}$&$0.289_{-0.024(0.060)}^{+0.036(0.065)}$&$-19.288_{-0.018(0.051)}^{+0.029(0.061)}$\\
        
        &CC+$\rm{PPS}$+BAO$_{1}$&$2.938^{+0.143(0.417)}_{-1.235(2.471)}$&$70.130_{-0.370(1.189)}^{+0.947(2.076)}$&$0.321_{-0.011(0.028)}^{+0.010(0.023)}$&$-19.331_{-0.014(0.039)}^{+0.024(0.052)}$ \\   
        &CC+$\rm{PPS}$+BAO$_{2}$&$1.750_{-0.123(0.299)}^{+0.103(0.281)}$&$70.126_{-0.302(0.698)}^{+0.276(0.701)}$&$0.306_{-0.006(0.016)}^{+0.006(0.016)}$&$-19.343_{-0.011(0.025)}^{+0.010(0.024)}$\\

        \hline  
        \hline

\end{tabular}
\caption{Confidence intervals ($68\%$ and $95\%$ levels) for the $f(R)$ models (Starobinsky, Hu-Saswicki, Exponential and  Hyperbolic Tangent) obtained using the different groups of observables selected to perform the statistical analyses.
(Cosmic Chronometers (CC), SneIa Pantheon$^+$+SH0ES (PPS) Collaboration and two different datasets from  BAO (BAO$_1$ and BAO$_2$)). 
}
\label{tab:results1}
\end{table*}

\begin{figure}[!h]
    \centering
    \includegraphics[width=0.95\linewidth]{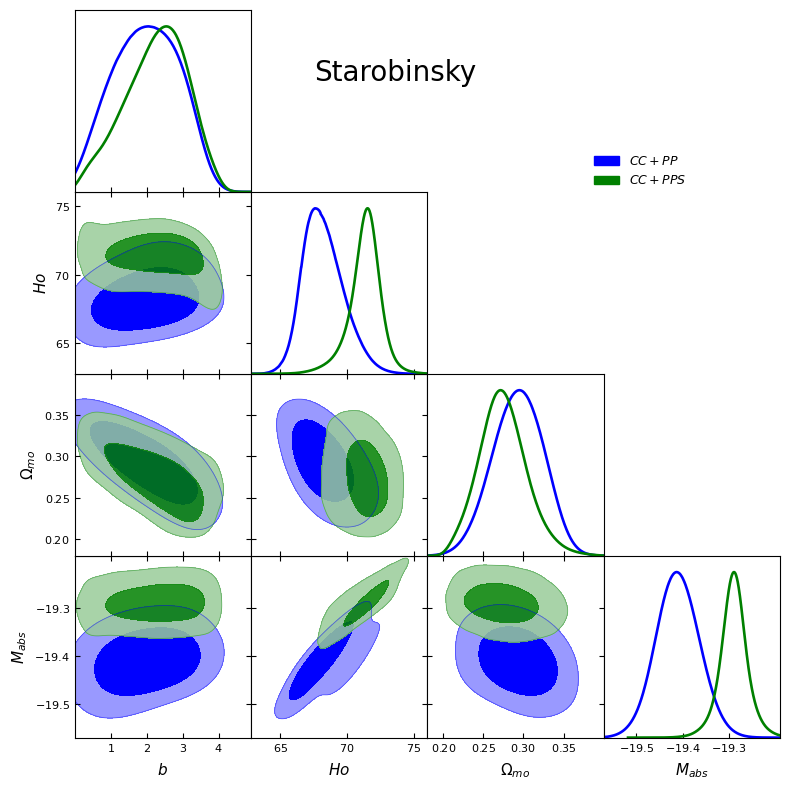}
    \caption{ Results of statistical analysis of $f(R)$ Starobinsky model  with and without using the cepheid-calibrated SnIa by SH0ES. The darker and brighter regions represent 1$\sigma$  (68 $\%$) and 2$\sigma$ (95$\%$) confidence intervals, respectively}
    \label{fig:staro0}
\end{figure}
            
\begin{figure}[!h]
    \centering
    \includegraphics[width=0.95\linewidth]{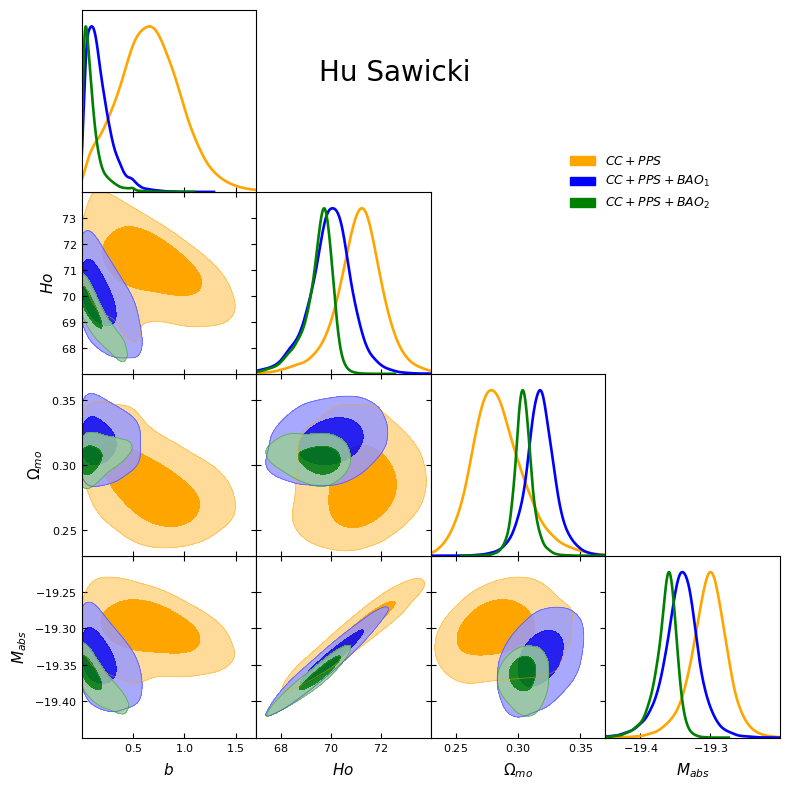}
    \caption{ Results of statistical analysis of $f(R)$ Hu Sawicki model  with different combinations of the data sets. The darker and brighter regions represent 1$\sigma$  (68 $\%$) and 2$\sigma$ (95$\%$) confidence intervals, respectively}
    \label{fig:hu}
\end{figure}

\begin{figure}[!h]
    \centering
    \includegraphics[width=0.95\linewidth]{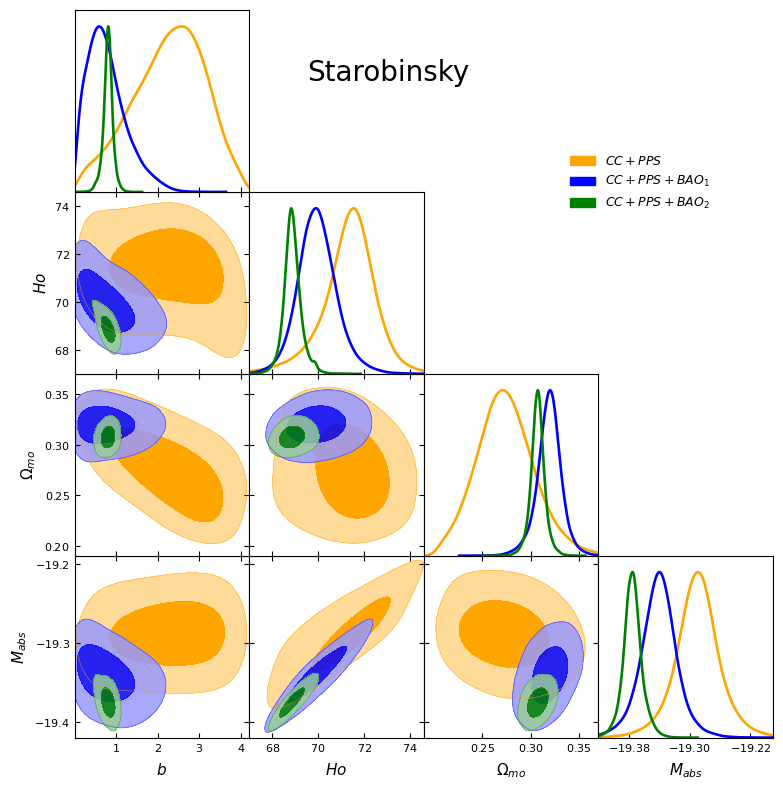}
    \caption{ Results of statistical analysis of $f(R)$ Starobinsky model  with different combinations of the data sets. The darker and brighter regions represent 1$\sigma$  (68 $\%$) and 2$\sigma$ (95$\%$) confidence intervals, respectively}
    \label{fig:staro}
\end{figure}

\begin{figure}[!h]
    \centering
    \includegraphics[width=0.95\linewidth]{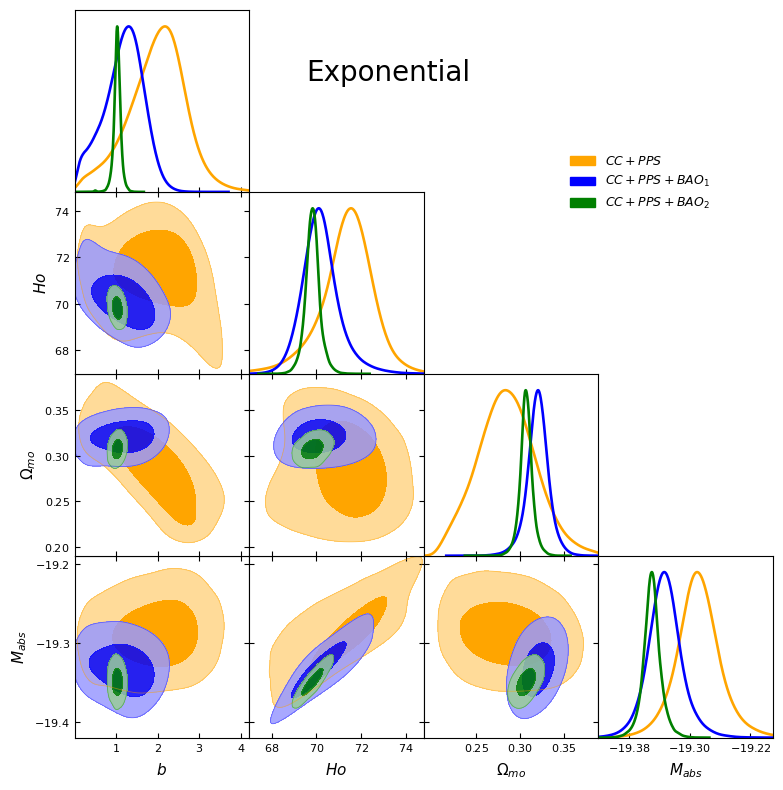}
    \caption{ Results of statistical analysis of $f(R)$ Exponential model  with different combinations of the data sets. The darker and brighter regions represent 1$\sigma$  (68 $\%$) and 2$\sigma$ (95$\%$) confidence intervals, respectively}
    \label{fig:exp}
\end{figure}
\begin{figure}[!h]
    \centering
    \includegraphics[width=0.95\linewidth]{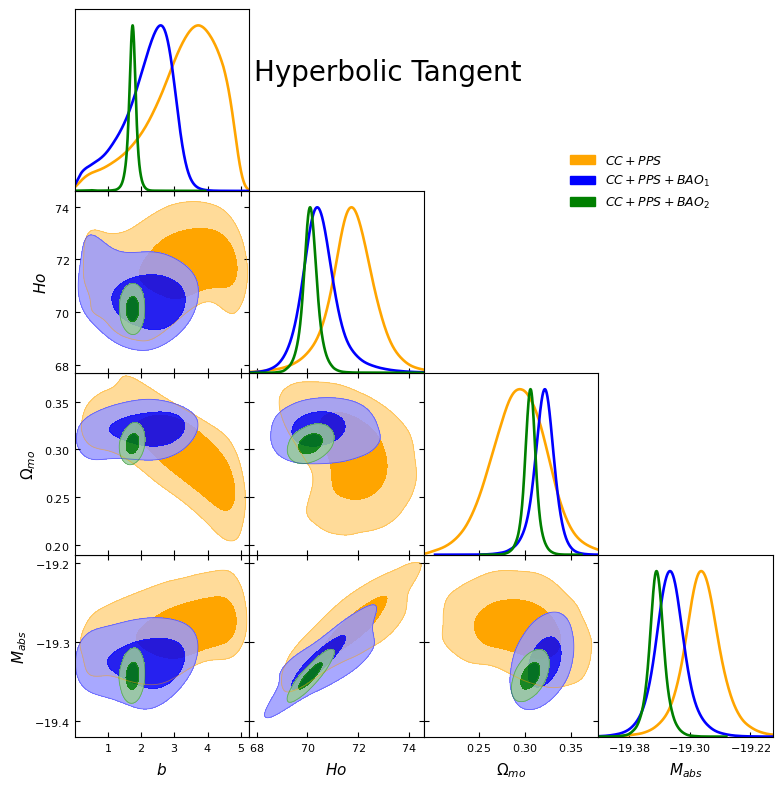}
    \caption{ Results of statistical analysis of $f(R)$ Hyperbolic Tangent model  with different combinations of the data sets. The darker and brighter regions represent 1$\sigma$  (68 $\%$) and 2$\sigma$ (95$\%$) confidence intervals, respectively}
    \label{fig:tanh}
\end{figure}

\subsection{The Hu-Sawicki model}

From the results presented in Table \ref{tab:results1} and Fig. \ref{fig:hu}, it is clear that in addition to further restricting the parameter spaces, both BAO data groups generate correlations between $b$ and $H_0$, and between $b$ and the $M_{abs}$ that are not as visible when only  CC and SnIa (PPS) are used. Also, the inclusion of these data reduces the degeneracies between the distortion parameter and $\Omega_{m0}$. On the other hand, the 1D posteriors of  parameters $b$, $H_0$ and $M_{abs}$ are shifted towards lower values, while for $\Omega_{m0}$ it is shifted towards higher values. This last shift is contrary to what is perceived in the standard cosmological model. 

On the other hand, for distortion parameter $b$,  the allowed zone is smaller than that obtained for the Starobinsky case (Fig. \ref{fig:all}), and closer to the null value (it is consistent with this value at 2$\sigma$ in the CC+PPS+BAO$_{1/2}$ analyses and almost also in the CC+PPS case). Since the Starobinsky model is equivalent to a Hu-Sawicki with $n=2$, it could be inferred that the higher the value of $n$, the greater the shift of the 1D probability posterior towards non-null values for $b$.

Furthemore, the results obtained for $\Omega_{m0}$, $H_0$ and $M_{abs}$ are consistent at 1$\sigma$ with those reported for the $\Lambda{\rm CDM}$ model for all analyses except in the CC+PPS case, where $\Omega_{m0}$ interval is compatible at 2$\sigma$; and in the CC+PPS+BAO$_2$ case, where the $H_{0}$ estimation is compatible at 2$\sigma$.\\
Finally, our findings using the BAO$_2$ dataset are significantly more constrained compared to those derived from the BAO$_1$ compilation,  indicating that the DESI DR2 data provides higher precision.

\subsection{The Starobinsky model}
\label{subsec:staro}
The behavior of this model (Fig. \ref{fig:staro}) closely parallels that of the previously studied case, especially regarding the constraining power of the BAO data (including the observed differences in limiting strength between the two BAO data sets), and to the regions obtained for the free parameters, the shifts of their 1D posterior distributions depending on the data sets used and the correlations between them with the exception of the distortion parameter $b$ (Fig. \ref{fig:all}). For this one, in all analyses, the estimated values are systematically shifted toward higher values, exhibiting 2$\sigma$-level inconsistencies with the null value, with the sole exception of the CC+PPS+BAO$_1$ combination, which remains consistent within 2$\sigma$.

Lastely, the results obtained for $\Omega_{m0}$, $H_0$ and $M_{abs}$ are consistent at 1$\sigma$ with those reported for the $\Lambda{\rm CDM}$ model for the
CC+PPS and CC+PPS+BAO$_1$ analyses but, for the CC+PPS case, $\Omega_{m0}$ interval is compatible at 2$\sigma$. However, for the CC+PPS+BAO$_2$ compilation, only the $\Omega_{m0}$ estimation is in agreement at 1$\sigma$, while $H_0$ and $M_{abs}$ are at 2$\sigma$.

\subsection{The Exponential model}

A clear difference between the above models and this one is that the estimated intervals for the distortion parameter $b$ are larger in the Exponential model (Fig \ref{fig:exp}). This is understandable since due to the functional form of this $f(R)$, a larger value of $b$ is needed in order to differentiate it from the $\Lambda{\rm CDM}$ model. However, the intervals obtained for the other free parameters are consistent with those of the previous models, perceiving the same types of correlation, 1D posterior distribution shifts and behaviors.

An interesting point to keep in mind is that although the results for parameter $b$ are not consistent with $\Lambda{\rm CDM}$ at 2$\sigma$ in any of the analyses, the estimated values for the other parameters are consistent at 1$\sigma$, with the exception of the value obtained for $H_0$ in the case CC+PPS+BAO$_2$, which is at 2$\sigma$.

\subsection{The Hyperbolic Tangent model}
As in the previous case, the functional form of this model (Fig. \ref{fig:tanh}) causes the intervals obtained for the distortion parameter to be shifted towards higher values. In addition, the shifts in the 1D posterior distributions and correlations of the parameters are very similar to those described for the other models, with the exception that, in this case, there is no defined correlation between the distortion parameter $b$ and $M_{abs}$, and between $b$ and $H_0$ as seen in the other $f(R)$ models.

Finally, as in the Exponential model, although the estimated interval for $b$ is not consistent with the null value at 2$\sigma$, there is consistency between the regions obtained for the other parameters and the regions calculated for $\Lambda{\rm CDM}$ at 1$\sigma$. 

\subsection{Model comparison}

We use different statistical criteria in order to determine which of the models presented in this article is  the
preferred one for the chosen data. The minimum value of the chi-square ($\chi^2$) is defined in relation to the maximum likelihood as $\chi^2_{min}= -2 \ln \mathcal{L}_{max}$, while $\chi^2_\nu=\chi^2_{min}/\nu$
is the reduced chi-square being $\nu=N-k$ the number of degrees of freedom, $N$ the total number of data points and $k$ the number of free parameters of the model. We also take into account the Akaike (AIC) \cite{1974ITAC...19..716A}  and Bayesian (BIC) \cite{10.1214/aos/1176344136}, \cite{doi:10.1177/0049124104268644}
information criterions that are given by
\begin{equation}
   { \rm AIC}=-2 \ln \mathcal{L}_{max}+2k
\end{equation}
\begin{equation}
   { \rm BIC}=-2 \ln \mathcal{L}_{max}+k\ln N 
\end{equation}
It is assumed that the model with the lowest values of $\chi^2_{min}$, AIC and BIC, is the one that best fits the data statistically. In contrast, in the case of reduced chi-square, the model most favored by the data is the one with an $\chi^2_{\nu}$ value closest to 1. If this value is $>>1$, the model is considered to be poor and therefore is subfitted while if it is $<<1$, the model is overfitted either because it is not adequate to fit for noise, or the error variance is overestimated \cite{Bevington1969}. Among these statistical criteria, the $\chi^2_{min}$ is the only one that does not consider the number of free parameters of the model to be analyzed. This point is very important since for a nested model, the greater the number of parameters, the greater the likelihood, without taking into account the relevance of the new parameters included. Conversely, the other criteria do penalize the models that include more free parameters and they do so in different ways, with the BIC being the most severe of the three. This sometimes leads to the criteria having discrepancies between them and it is necessary to analyze whether the assumptions of the criteria are being violated (\cite{doi:10.1177/0049124104268644}, \cite{2004MNRAS.351L..49L} and many others).
On the other hand, the comparison of criteria of two models  is carried out by subtracting the values obtained by both models ($\Delta X=\Delta AIC$ or $\Delta X=\Delta BIC$) such that: a) if $0\le\Delta X \le2$ or $-2\le\Delta X <0$ the evidence is weak, it cannot be defined which is better; b) if $2<\Delta X \le6$ or $-6\le\Delta X <-2$ the evidence is positive; c) if $6<\Delta X \le10$ or $-10\le\Delta X <6$ the evidence is strong; and d) if $\Delta X>10 $ or $\Delta X <-10$ the evidence is very strong.\\
From the results shown in Table \ref{tab:model_comparison} and the description above, it can be seen that the data combination CC+PPS+BAO$_1$ generates higher likelihoods for the models analyzed than the combination CC+PPS+BAO$_2$. For the 
the Hu-Sawicki model, only for the CC+PPS+BAO$_1$ data set the $\chi^2_{min}$ is lower compared to the one from $\Lambda{\rm CDM}$. For both data combinations, the reduced chi-squared values ($\chi^2_{\nu}$) are closer to unity; however, the AIC and BIC values are higher, which leads to evidence favoring the $\Lambda{\rm CDM}$ model over the Hu-Sawicki model-weakly and strongly, respectively when the BAO$_1$ combination is used, and strongly for both criteria in the other case.

Meanwhile, for Starobinsky model for both data combinations is given that the $\chi^2_{min}$ are lower than those of $\Lambda{\rm CDM}$. However, only the $\chi^2_{\nu}$ obtained with the CC+PPS+BAO$_1$ dataset is closer to one.

On the other hand, based on the analysis of Bayesian information criteria, there are weak (AIC) and strong (BIC) evidences in favor of the standard cosmological model over the Starobinsky model for the CC+PPS+BAO$_1$ dataset. In contrast, for the other case ($CC+PPS+BAO_2$), there is very strong preference for the $f(R)$ model according to both the $\Delta$AIC and $\Delta$BIC. This result is remarkable, as despite having an additional free parameter, which is primarily penalized
by the BIC index, it remains statistically preferred.

Finally from the values obtained for the  the Exponential and Hyperbolic Tangent models it follows that for both data compilations the $\chi^2_{min}$ reach lower values than those of $\Lambda{\rm CDM}$ as well as lower AIC values, obtaining weak evidence supporting these $f(R)$ for the case CC+PPS+BAO$_1$ and furthemore, very strong evidence for the case CC+PPS+BAO$_2$. However, this does not occur when both, the results for $\chi^2_{\nu}$  and BIC, are analyzed. The  $\chi^2_{\nu}$ values are less closer to one and the BIC criterias are higher such that the evidence benefiting $\Lambda{\rm CDM}$ is positive when CC+PPS+BAO$_1$ is used. Notably, for the CC+PPS+BAO$_2$ case we obtain very strong evidence in favor of these last $f (R)$ models in comparison with the $\Lambda$CDM model.

\begin{table*}
    \centering
    \begin{tabular}{lccccccc}
        \hline
        \hline
      Data&  Model       & $\chi^2_{min}$ & $\chi^2_{\nu}$ & AIC & BIC & $\Delta$AIC & $\Delta$BIC \\
        \hline
       CC+PPS+BAO$_1$& $\Lambda$CDM   & 1536.492 & 0.9615 & 1542.492 & 1558.627 & - & - \\
        &Hu-Sawicki   & 1536.373 & 0.9620 & 1544.373 & 1565.887 & 1.881 & 7.259 \\
        &Starobinsky    & 1535.791 & 0.9617 & 1543.791 & 1565.304 & 1.299 & 6.677 \\
        &Exponential   & 1534.157 & 0.9606 & 1542.157 & 1563.671 & -0.334 & 5.043 \\
        &Hyperbolic Tangent  & 1533.201 & 0.9600 & 1541.201 &  1562.715 &  -1.290 & 4.088 \\
        \hline
        CC+PPS+BAO$_2$& $\Lambda$CDM   & 1585.536 & 0.9272 & 1591.536 & 1607.874 & - & - \\
        & Hu-Sawicki     & 1588.178 & 0.9293 & 1596.178 & 1617.962 & 4.642 & 10.088 \\
        & Starobinsky    & 1563.613 & 0.9149 & 1571.613 & 1593.397 & -19.922 & -14.476 \\
        & Exponential    & 1556.196 & 0.9106 & 1564.196 & 1585.980 & -27.340 & -21.894 \\
        & Hyperbolic Tangent  & 1557.191 & 0.9112 & 1565.191 & 1586.975 & -26.345 & -20.899 \\
        \hline
        \hline
    \end{tabular}
    \caption{Results of standard statistical tools commonly used to assess model fitting: the chi-square and the  reduced chi-square statistics ($ \chi^2_{min}$, $ \chi^2_\nu$), the Akaike Information Criterion (AIC), and the Bayesian Information Criterion (BIC) for each model.}
    \label{tab:model_comparison}
\end{table*}
\begin{figure}[H]
    \centering
    \includegraphics[width=0.9\linewidth]{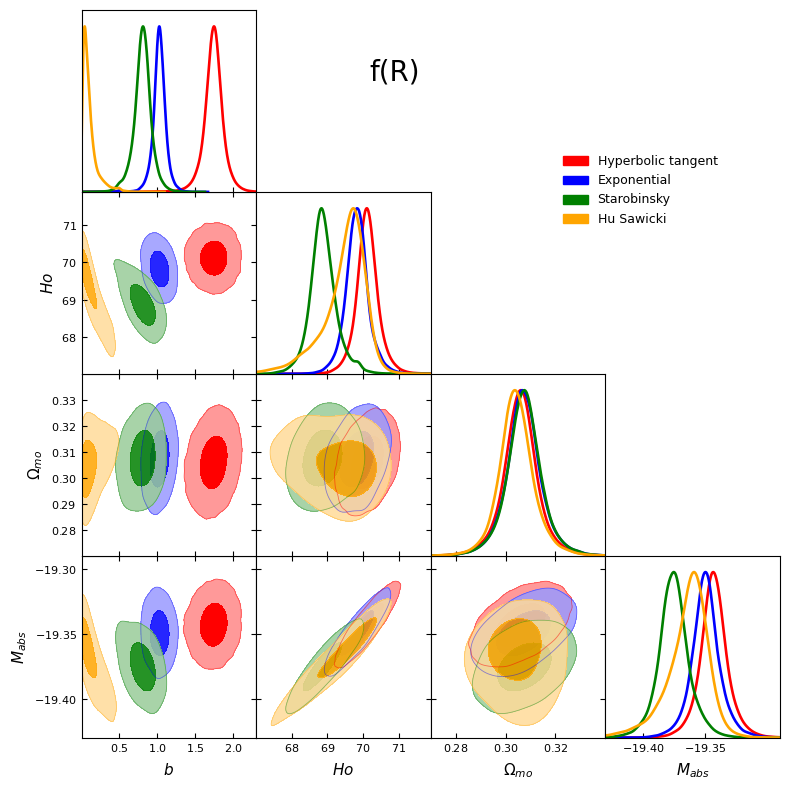}
    \caption{Results for the distortion parameter $b$, the Hubble parameter $H_0$, the matter density parameter $\Omega_{m0}$ and the absolute magnitude $M_{abs}$ for all the $f(R)$ models studied in this work obtained using combinations of the Cosmic Chronometers, Type Ia Supernovae data from the Pantheon$^+$+SH0ES collaborations, and the new DESI$_{DR2}$ BAO data. The plots show the  $68\%$ and $95\%$ confidence regions.}
    \label{fig:all}
\end{figure}

\section{Discussion}
\label{Sec:discussion}

In this section we compare our results with those published by other authors for the same $f(R)$ models with the same data but different developments and with similar data. 
Our analysis shows that the use of data from the Pantheon$^+$ compilation increases the size and shifts the 1D distribution towards higher values of the distortion parameter $b$. This can also be seen when comparing our results with those obtained in 
\cite{ODINTSOV2021115377,M2022PhRvD.105j3526L,furrugia2021PhRvD.104l3503F} (among many others) where Pantheon data were used and with those published in \cite{2024MNRAS.527.7626R,PhysRevD.109.123514,2024arXiv241209409O} where the previous data were replaced by Pantheon$^+$. However, this effect is not reflected in \cite{Kumar:2023bqj}, but this is because in said analysis only the approximations developed by Basilakos \cite{PhysRevD.87.123529} were used to obtain the expressions of $H(z)$ for the Hu-Sawicki and Starobinsky models. 
This procedure, being a perturbative method around $\Lambda{\rm CDM}$, is not appropriate for exploring regions of the distortion parameter $b$ that are far from $b = 0$ (corresponding to $\Lambda{\rm CDM}$) such as those presented when using Pantheon$^+$ compilation. On the other hand, the addition of cepheid-calibrated supernovae by the SH0ES collaboration to the Pantheon$^+$ compilation results in estimated values for $H_0$ and $M_{abs}$ being higher than those estimated when using Pantheon$^+$ data alone.

Our estimates for the Starobinsky and Hu-Sawicki models using CC+PPS data are  consistent at 1$\sigma$ with those published in \cite{PhysRevD.109.123514} for the CC+PPM\footnote{PPM refers to the Pantheon$^+$ data compilation with a marginalization on the $M_{abs}$ proposed by Camarena $\&$ Marra \cite{Camarena2023}} data sets. The small differences between the intervals found may be due to the fact that in one case calibrated supernovae with Cepheids by the SH0ES collaboration were used (PPS compilation) while in the other an statistical
marginalization on the $M_{abs}$ obtained from the SH0ES data was used (PPM). 

There is also almost\footnote{The confidence intervals for $\Omega_m$ in the exponential model, and for both $\Omega_m$ and $b$ in the hyperbolic tangent model, derived using the CC+PPS+BAO$_2$ dataset, show slight deviations from those reported by previous authors} 1$\sigma$  consistency with the results reported by \cite{2024MNRAS.527.7626R} for the four $f(R)$ models using PPS+CC, PPS+CC+BAO$_1$ and PPS+CC+BAO$_2$ data sets, although the BAO data points are not exactly the same (in addition to not containing the DESI data). This paper also uses in some of ther analysis data from HII regions of galaxies, and parameters obtained from CMB data, such as the acoustic scale $l^{Pl}_A$, the displacement parameter $R^{Pl}_A$ and the present baryon density $\omega_b^{Pl}=\frac{1}{4}\Omega_bh^2$. The latter, are estimated through a statistical analysis where a $\Lambda{\rm CDM}$ is taken as a fiducial model. So, like the authors of \cite{M2022PhRvD.105j3526L}, we consider that 
despite being a datum that has a greater restrictive power than BAO,
its use for testing alternative cosmological models is not the most appropriate \footnote{In addition, another issue to consider is
that if the $H_0$ estimate obtained with CMB data (without using $H_0$ estimates from SH0ES and/or no marginalization on the $M_{abs}$ from said collaboration) is not consistent with the result published by SH0ES Collaboration, it is not correct to carry out a statistical study incorporating both data at the same time.}. Furthermore, it is not clear whether the authors use the entire data compilation provided by Pantheon$^+$ or exclude those with $z<0.01$ where the sensitivity of the SnIa peculiar velocities is very large and so are their uncertainties (details in \cite{2022ApJ...938..110B,10.1093/mnras/stad451} and a private communication with D. Brout). This last issue is also reflected in the analysis of the Exponential model performed by Odinstov et al.\cite{2024arXiv241209409O} where they also use the Pantheon$^+$ data (without SH0ES data or other $H_0$ estimates or marginalizations on $M_{abs}$), cosmic chronometers and CMB data described above. In addition, they incorporate the DESI DR1 data to previous BAO observations, but this should be done taking into account the suggestions explained in the DESI collaboration paper \cite{2024arXiv240403002D} that were ignored.
Only their $\Omega_{m0}$ result is consistent
at 1$\sigma$ with ours. This is mainly due to the use of SH0ES and CMB data \footnote{The first data set  generates a shift of the 1D posterior distribution of $H_0$ towards higher values, and a slight shift towards lower values in the case of parameter $b$; while the second, triggers a slight shift of the 1D posterior distribution of $b$ towards higher values, and a shift towards lower values of $H_0$.}, since when comparing the estimates reported by \cite{2024MNRAS.527.7626R} without SH0ES with those of \cite{2024arXiv240403002D} there is also consistency between the rest of the parameters.

\section{Conclusions}
\label{Sec:conclusions}

\par In this article, we have analyzed four of the most widely studied $f(R)$ gravity models—namely, the Hu-Sawicki, Starobinsky, Exponential, and Hyperbolic Tangent models—within a cosmological framework, using the latest compilation of Baryon Acoustic Oscillation (BAO) measurements from the DESI DR2 release \citep{DR22025arXiv250314738D}.

In addition, we employed the most recent Type Ia supernova data from the Pantheon+ compilation \citep{2022ApJ...938..110B,2022ApJ...938..113S}, along with data that belong from 42 SnIa calibrated via Cepheid variables, as analyzed by the SH0ES collaboration \citep{2022ApJ...938..110B,2022ApJ...938..113S}. The latter dataset allows us to avoid marginalization over the absolute magnitude $M_{\rm abs}$. Complementary data from cosmic chronometers (CC) and previous BAO compilations were also included, as shown in Table \ref{tab:BAO_data}, to investigate the improvement in constraint capabilities provided by the new DESI DR2 BAO dataset.

To achieve this study, we solved the modified Friedmann equations while executing a Markov Chain Monte Carlo (MCMC) statistical method at each step, using the aforementioned observational datasets. The final model comparison was performed using Bayesian statistical criteria, with the standard $\Lambda \text{CDM}$ model taken as the reference.

With regard to the comparison between $\text{BAO}_1$ and $\text{BAO}_2$ data, our results reaffirm those presented in \citep{2024arXiv240403002D,DR22025arXiv250314738D}, highlighting the remarkable increase in constraining power achieved by the new measurements.

Analyzing the confidence intervals for the distortion parameter $b$, we find that when using $\text{CC} + \text{PPS} + \text{BAO}_1$, the Hu-Sawicki and Starobinsky models are incompatible with the standard model at the $1\sigma$ level, while the Exponential and Hyperbolic Tangent models are incompatible at more than $2\sigma$. This result changes when $\text{BAO}_1$ is replaced by $\text{BAO}_2$: in this case, only the Hu-Sawicki model remains inconsistent at the $1\sigma$ level, and the other three models show inconsistencies exceeding $2\sigma$.
Additionally, we observe that for both the Hu-Sawicki and Starobinsky models—the latter being equivalent to the Hu-Sawicki model with $n = 2$—larger values of the parameter $n$ consistently result in higher best-fit values of the distortion parameter $b$, independent of the BAO dataset used. In all cases, the use of $\text{BAO}_2$ data leads to lower estimated values of the Hubble parameter $H_0$ compared to results obtained with $\text{BAO}_1$. It is also observed that all $f(R)$ models yield $H_0$ values lower than those predicted by the standard model. As for the total matter density parameter $\Omega_m$, all models are mutually consistent and compatible with the standard $\Lambda \text{CDM}$ model.
Finally, when analyzing the Bayesian statistical indicators, we find that for the $\text{BAO}_1$ dataset, the comparisons with $\Lambda \text{CDM}$ yield weak evidence or statistical indistinguishability according to the $\Delta \text{AIC}$ index in all four cases. According to the $\Delta \text{BIC}$ index, strong evidence is obtained in favor of the standard model for the Hu-Sawicki and Starobinsky models, and positive evidence for the Exponential and Hyperbolic Tangent models. The preference for $\Lambda \text{CDM}$ indicated by $\Delta \text{BIC}$ is due to the penalization of the additional free parameter present in the alternative models.
When the analysis is repeated with $\text{BAO}_2$ instead of $\text{BAO}_1$, the conclusions change drastically. The Starobinsky, Exponential, and Hyperbolic Tangent models all show very strong evidence in their favor over the standard $\Lambda \text{CDM}$ model. This unprecedented result is attributed to the high precision of the new DESI DR2 data \citep{DR22025arXiv250314738D}, which, for the first time, reveals a statistical preference for the alternative models studied herein, as compared to the standard scenario. This outcome is of great relevance for future investigations into gravitational physics beyond General Relativity.
The fact that all models except for Hu-Sawicki obtain very strong evidence in contrast to the standard model is reasonable when considering the aforementioned behavior: the models that show greater deviation from $\Lambda \text{CDM}$ are precisely those statistically preferred. The Hu-Sawicki model, by construction, allows minimal deviation of the distortion parameter $b$ from zero. As previously stated, larger values of $n_{\text{HS}}$ permit more flexibility in $b$, and the specific case analyzed here corresponds to the minimal value $(n = 1)$. Therefore, a model that deviates little from $\Lambda \text{CDM}$ and introduces an additional free parameter is not expected to show statistical preference, particularly under $\Delta \text{BIC}$.

Concerning the $H_0$ tension, although the $f(R)$ models analyzed here are statistically preferred over $\Lambda \text{CDM}$, they do not appear to alleviate the discrepancy \citep{2016MNRAS.459.3880H,planckAghanim:2018eyx,
riess2022ApJ...934L...7R,2024A&A...686A.193K}. However, future developments—such as the use of alternative calibration methods for Type Ia supernovae \citep{2024arXiv240806153F,peri2024PhRvD.110l3518P}—may help alleviate the tension by altering the model-independent determination of $H_0$. At present, though, no definitive claims can be made.

In conclusion, we highlight the substantial change in the outcomes of the Bayesian model comparisons, driven exclusively by the enhanced constraining power of the new $\text{BAO}_2$ (DESI DR2) dataset, which allows for more rigorous model discrimination. We therefore conclude that $f(R)$ gravity remains a compelling theoretical framework for explaining the late-time accelerated expansion of the universe, now supported by statistical preference over $\Lambda \text{CDM}$ in light of the most recent observations.

The fact that the Starobinsky $f(R)$ model (with an $R^2$ term), which provides the best fit to the Cosmic Microwave Background data in the early universe and serves as the reference model in Bayesian analyses in that regime \citep{planckinflation2020A&A...641A..10P}, now finds complementary support through late-time statistical preference for $f(R)$ models as reported here, offers a promising outlook for continued exploration of this class of modified gravity theories.

\section{Acknowledgments}
The authors would like to thank D. Brout, M. Leizerovich, A. Ballatore and S. J. Landau for their helpful comments.


The authors are supported by CONICET Grant No. PIP 11220200100729CO, and Grants No.G175 from UNLP.
\par 

\appendix
\section{Methods for solving the Friedmann equations in specific $f(R)$ models}
\label{Sec:Methods}

The solution of the Friedmann equations \ref{eq:friedmod} often requires numerical integration, as analytical solutions are not generally feasible. To enhance computational efficiency and mitigate numerical instabilities, it is advantageous to tailor the methodology to the specific characteristics of each $f(R)$ model. Below, we outline the approaches used for the models under consideration.

\subsection{Starobinsky and Hu-Sawicki Models}

In order to optimize computational efficiency when solving the Friedmann equations, for the Hu-Sawicki and Starobinsky models we use the variable transformation proposed by de la Cruz-Dombriz \textit{et al.} \cite{2016PhRvD..93h4016D}. The set of variables is defined as: 

\begin{subequations}
\begin{align}
x & = \frac{\dot{R} f_{RR}}{H f_{R}}, \\
y & = \frac{f}{6H^2 f_{R}}, \\
v & = \frac{R}{6H^2}, \\
\Omega & = \frac{8\pi G \rho_m}{3H^2 f_{R}}, \\
\Gamma & = \frac{f_R}{R f_{RR}}, \\
r & = R / R^*,
\end{align}
\end{subequations}
where $R^*=R_{HS},R_{S}$ is a reference curvature scale. Using these variables, the field equations take the form:
\begin{subequations}
\begin{align}
\frac{{\rm d} H}{{\rm d} z} & = \frac{H}{z+1}\left(2 - v\right), \\
\frac{{\rm d} x}{{\rm d} z} & = \frac{1}{z+1}\left(-\Omega - 2v + x + 4y + xv + x^2\right), \\
\frac{{\rm d} y}{{\rm d} z} & = \frac{-1}{z+1}\left(vx\Gamma - xy + 4y - 2yv\right), \\
\frac{{\rm d} v}{{\rm d} z} & = \frac{-v}{z+1}\left(x\Gamma + 4 - 2v\right), \\
\frac{{\rm d} \Omega}{{\rm d} z} & = \frac{\Omega}{z+1}\left(-1 + 2v + x\right), \\
\frac{{\rm d} r}{{\rm d} z} & = -\frac{x\Gamma r}{z+1}.
\end{align}
\end{subequations}
The redshift associated with the initial conditions for all models will correspond to the moment when the model approaches the $\Lambda{\rm CDM}$ model closely enough, with which these conditions are calculated. This can be expressed as:
\[
f(R(z_i)) = R-2\Lambda (1 - \epsilon)
\]
With this condition, the following expression is derived \citep{furrugia2021PhRvD.104l3503F}:
\begin{equation}
z_i =
\left[ \frac{4 \Omega^{\Lambda}_{\Lambda,0}}{\Omega^{\Lambda}_{m_0}}\left(\frac{b}{4 \nu_{f(R)}} - 1\right)\right]^{1/3}-1,
\label{eq:zi}
\end{equation}
Where the expression for $\nu_{f(R)}$ in each model, as well as the assumed values for \( \epsilon \) \citep{furrugia2021PhRvD.104l3503F}, are:

\begin{align}
\nu_{hs} &= \frac{1}{(1/\epsilon - 1)},\  \epsilon\approx  10^{-6}  \\
\nu_{st} &= \left(\frac{1}{(1/\epsilon - 1)} \right)^{1/2},\  \epsilon\approx  10^{-8} \\
\nu_{exp} &= -\frac{1}{\ln \epsilon},\  \epsilon\approx  10^{-10} \\
\nu_{tanh} &= \frac{1}{\operatorname{arctanh}(1 - \epsilon)},\  \epsilon\approx  10^{-10} \\
\end{align}
\label{eq:nu_f}
Therefore, the initial conditions for Starobinsky and Hu Sawicki cases are given by:

\begin{align}
    x_i &= 0  \\
    y_i &= \frac{R_{\Lambda\text{CDM}}(z_i) - 2\Lambda}{6H^2_{\Lambda\text{CDM}}(z_i)}   \\
    v_i &= \frac{R_{\Lambda\text{CDM}}(z_i)}{6H^2_{\Lambda\text{CDM}}(z_i)}   \\
    \Omega_i &= 1 - v_i + x_i + y_i  \\
    r_i &= \frac{R_{\Lambda\text{CDM}}(z_i)}{R_{j}} 
\end{align}
where \( R_{j} = R_{S} \) or \( R_{HU} \), depending on the case. \( R_{\Lambda CDM} \) and \( H_{\Lambda CDM} \) are the Ricci scalar and the Hubble function of the standard model, respectively.

Since the \( f(R) \) model reduces to the standard model when \( b = 0 \), one can assume, as a solution for \( H(z) \), a series expansion around the standard model solution, proposed by Basilakos \textit{et al.} \cite{PhysRevD.87.123529}, valid only for small \( b \), yielding a very low error. 

\begin{equation}
\label{eq:exp-ser}
H^2(N) = H_{\Lambda {\rm CDM}}^2(N) + \sum_{i=1}^M b^i \delta H_i^2(N),
\end{equation}

where $N = -\log(1+z)$.

This approach avoids numerical instabilities associated with $f_{RR} \to 0$. For cases with \( b \) smaller than a critical value \( b_{\text{crit}} \), this solution is assumed, with a negligible error as indicated in \citep{PhysRevD.87.123529}.
For larger \( b \), the full modified Friedmann equations \ref{eq:friedmod} and \ref{eq:friedmod2} are solved numerically.
\subsection{Exponential and Hyperbolic Tangent Models}
For the exponential and hyperbolic tangent models meanwhile, we employ a different approach based on a redefinition of variables, which is presented in \cite{Odintsov2017}. This method reformulates the Friedmann equations in terms of dimensionless variables, expressed as:

\begin{subequations}
\begin{align}
\label{eq:motionExp1}
\frac{{\rm d} E}{{\rm d} x} & ={\Omega^{\Lambda {\rm CDM}}_{\Lambda,0}}\frac{\mathcal{R}}{E}-2E, \\
\label{eq:motionExp2}
\frac{{\rm d} \mathcal{R}}{{\rm d} x} & =\frac{2\Lambda}{f_{\mathcal{R}\mathcal{R}}} \Bigg[\Omega^{\Lambda {\rm CDM}}_{m,0} \frac{a^{-3} + X^{\Lambda {\rm CDM}}a^{-4}}{E^2} - \frac{f_{\mathcal{R}}}{2\Lambda} \notag \\
&+ \frac{\mathcal{R}f_{\mathcal{R}} - f}{6\left(H_0^{\Lambda {\rm CDM}}\right)^2E^2} \Bigg],
\end{align}
\end{subequations}

where $E = H / H_0^{\Lambda {\rm CDM}}$,$X^{\Lambda CDM} = \frac{\Omega^{\Lambda CDM}_{r,0}}{\Omega^{\Lambda CDM}_{m,0}}$,\\ $\mathcal{R} = R / (2\Lambda)$, and $x = -\log(1+z)$. \( f_{\mathcal{R}} \) and \( f_{\mathcal{RR}} \) are the first and second derivatives with respect to \( \mathcal{R} \).

The initial conditions for the exponential and hyperbolic tangent cases are given by:

\begin{subequations}
\begin{align}
E^2(x_i) & = \Omega^{\Lambda {\rm CDM}}_{m,0} \left(e^{-3x_i} + X^{\Lambda {\rm CDM}} e^{-4x_i} \right) + \Omega^{\Lambda {\rm CDM}}_{\Lambda,0}, \\
\mathcal{R}(x_i) & = 2 + \frac{\Omega^{\Lambda {\rm CDM}}_{m,0}}{2\Omega^{\Lambda {\rm CDM}}_{\Lambda,0}} e^{-3x_i},
\end{align}
\end{subequations}
In most cases, the redshift calculated for the initial condition $z_i$ is greater than the highest redshift in the observations, which we will denote as $z_{\text{max}}$.
For intermediate cases, where a value of \( b \) results in \( z_i < z_{\text{max}} \), the \( f(R) \) model solution obtained from Eqs. \ref{eq:motionExp1} and \ref{eq:motionExp2} is assumed when \( z < z_i \); while the standard solution $\Lambda_{CDM}$ is considered in cases where \( z_i< z \).

For parameter sets with extremely small \( b \), \( z_i \) (Eq. \ref{eq:zi}) takes negative values, indicating that the model behaves indistinguishably from the standard model for all redshifts. Therefore, we assume $\Lambda_{CDM}$ as a good approximation in such cases.
This is very convenient since in the Exponential and Hyperbolic Tangent models, it is not possible to use a series expansion around the standard solution \ref{eq:exp-ser} as in the Starobinsky and Hu-Sawicki models, and in this methodology, the divergence \( f_{RR} \to 0 \) is also avoided.

\newpage
\bibliographystyle{apsrev}
\bibliography{testmog}

\begin{thebibliography}{79}
\expandafter\ifx\csname natexlab\endcsname\relax\def\natexlab#1{#1}\fi
\expandafter\ifx\csname bibnamefont\endcsname\relax
  \def\bibnamefont#1{#1}\fi
\expandafter\ifx\csname bibfnamefont\endcsname\relax
  \def\bibfnamefont#1{#1}\fi
\expandafter\ifx\csname citenamefont\endcsname\relax
  \def\citenamefont#1{#1}\fi
\expandafter\ifx\csname url\endcsname\relax
  \def\url#1{\texttt{#1}}\fi
\expandafter\ifx\csname urlprefix\endcsname\relax\def\urlprefix{URL }\fi
\providecommand{\bibinfo}[2]{#2}
\providecommand{\eprint}[2][]{\url{#2}}

\bibitem[{\citenamefont{{Riess} et~al.}(1998)\citenamefont{{Riess}, {Filippenko}, {Challis}, {Clocchiatti}, {Diercks}, {Garnavich}, {Gilliland}, {Hogan}, {Jha}, {Kirshner} et~al.}}]{rii1998AJ....116.1009R}
\bibinfo{author}{\bibfnamefont{A.~G.} \bibnamefont{{Riess}}}, \bibinfo{author}{\bibfnamefont{A.~V.} \bibnamefont{{Filippenko}}}, \bibinfo{author}{\bibfnamefont{P.}~\bibnamefont{{Challis}}}, \bibinfo{author}{\bibfnamefont{A.}~\bibnamefont{{Clocchiatti}}}, \bibinfo{author}{\bibfnamefont{A.}~\bibnamefont{{Diercks}}}, \bibinfo{author}{\bibfnamefont{P.~M.} \bibnamefont{{Garnavich}}}, \bibinfo{author}{\bibfnamefont{R.~L.} \bibnamefont{{Gilliland}}}, \bibinfo{author}{\bibfnamefont{C.~J.} \bibnamefont{{Hogan}}}, \bibinfo{author}{\bibfnamefont{S.}~\bibnamefont{{Jha}}}, \bibinfo{author}{\bibfnamefont{R.~P.} \bibnamefont{{Kirshner}}}, \bibnamefont{et~al.}, \bibinfo{journal}{\aj} \textbf{\bibinfo{volume}{116}}, \bibinfo{pages}{1009} (\bibinfo{year}{1998}), \eprint{astro-ph/9805201}.

\bibitem[{\citenamefont{{Perlmutter} et~al.}(1999)\citenamefont{{Perlmutter}, {Aldering}, {Goldhaber}, {Knop}, {Nugent}, {Castro}, {Deustua}, {Fabbro}, {Goobar}, {Groom} et~al.}}]{Perlmutter1999ApJ...517..565P}
\bibinfo{author}{\bibfnamefont{S.}~\bibnamefont{{Perlmutter}}}, \bibinfo{author}{\bibfnamefont{G.}~\bibnamefont{{Aldering}}}, \bibinfo{author}{\bibfnamefont{G.}~\bibnamefont{{Goldhaber}}}, \bibinfo{author}{\bibfnamefont{R.~A.} \bibnamefont{{Knop}}}, \bibinfo{author}{\bibfnamefont{P.}~\bibnamefont{{Nugent}}}, \bibinfo{author}{\bibfnamefont{P.~G.} \bibnamefont{{Castro}}}, \bibinfo{author}{\bibfnamefont{S.}~\bibnamefont{{Deustua}}}, \bibinfo{author}{\bibfnamefont{S.}~\bibnamefont{{Fabbro}}}, \bibinfo{author}{\bibfnamefont{A.}~\bibnamefont{{Goobar}}}, \bibinfo{author}{\bibfnamefont{D.~E.} \bibnamefont{{Groom}}}, \bibnamefont{et~al.}, \bibinfo{journal}{\apj} \textbf{\bibinfo{volume}{517}}, \bibinfo{pages}{565} (\bibinfo{year}{1999}), \eprint{astro-ph/9812133}.

\bibitem[{\citenamefont{{Riess} and {Breuval}}(2024)}]{3riess2024IAUS..376...15R}
\bibinfo{author}{\bibfnamefont{A.~G.} \bibnamefont{{Riess}}} \bibnamefont{and} \bibinfo{author}{\bibfnamefont{L.}~\bibnamefont{{Breuval}}}, in \emph{\bibinfo{booktitle}{IAU Symposium}}, edited by \bibinfo{editor}{\bibfnamefont{R.}~\bibnamefont{{de Grijs}}}, \bibinfo{editor}{\bibfnamefont{P.~A.} \bibnamefont{{Whitelock}}}, \bibnamefont{and} \bibinfo{editor}{\bibfnamefont{M.}~\bibnamefont{{Catelan}}} (\bibinfo{year}{2024}), vol. \bibinfo{volume}{376} of \emph{\bibinfo{series}{IAU Symposium}}, pp. \bibinfo{pages}{15--29}, \eprint{2308.10954}.

\bibitem[{\citenamefont{{Planck Collaboration} et~al.}(2020{\natexlab{a}})\citenamefont{{Planck Collaboration}, {Aghanim}, {Akrami}, {Ashdown}, {Aumont}, {Baccigalupi}, {Ballardini}, {Banday}, {Barreiro}, {Bartolo} et~al.}}]{Planckcosmo2018}
\bibinfo{author}{\bibnamefont{{Planck Collaboration}}}, \bibinfo{author}{\bibfnamefont{N.}~\bibnamefont{{Aghanim}}}, \bibinfo{author}{\bibfnamefont{Y.}~\bibnamefont{{Akrami}}}, \bibinfo{author}{\bibfnamefont{M.}~\bibnamefont{{Ashdown}}}, \bibinfo{author}{\bibfnamefont{J.}~\bibnamefont{{Aumont}}}, \bibinfo{author}{\bibfnamefont{C.}~\bibnamefont{{Baccigalupi}}}, \bibinfo{author}{\bibfnamefont{M.}~\bibnamefont{{Ballardini}}}, \bibinfo{author}{\bibfnamefont{A.~J.} \bibnamefont{{Banday}}}, \bibinfo{author}{\bibfnamefont{R.~B.} \bibnamefont{{Barreiro}}}, \bibinfo{author}{\bibfnamefont{N.}~\bibnamefont{{Bartolo}}}, \bibnamefont{et~al.}, \bibinfo{journal}{\aap} \textbf{\bibinfo{volume}{641}}, \bibinfo{eid}{A6} (\bibinfo{year}{2020}{\natexlab{a}}), \eprint{1807.06209}.

\bibitem[{\citenamefont{{Joyce} et~al.}(2016)\citenamefont{{Joyce}, {Lombriser}, and {Schmidt}}}]{joyce2016ARNPS..66...95J}
\bibinfo{author}{\bibfnamefont{A.}~\bibnamefont{{Joyce}}}, \bibinfo{author}{\bibfnamefont{L.}~\bibnamefont{{Lombriser}}}, \bibnamefont{and} \bibinfo{author}{\bibfnamefont{F.}~\bibnamefont{{Schmidt}}}, \bibinfo{journal}{Annual Review of Nuclear and Particle Science} \textbf{\bibinfo{volume}{66}}, \bibinfo{pages}{95} (\bibinfo{year}{2016}), \eprint{1601.06133}.

\bibitem[{\citenamefont{{Tsujikawa}}(2013)}]{suji2013CQGra..30u4003T}
\bibinfo{author}{\bibfnamefont{S.}~\bibnamefont{{Tsujikawa}}}, \bibinfo{journal}{Classical and Quantum Gravity} \textbf{\bibinfo{volume}{30}}, \bibinfo{eid}{214003} (\bibinfo{year}{2013}), \eprint{1304.1961}.

\bibitem[{\citenamefont{{Copeland} et~al.}(2006)\citenamefont{{Copeland}, {Sami}, and {Tsujikawa}}}]{2006copelandIJMPD..15.1753C}
\bibinfo{author}{\bibfnamefont{E.~J.} \bibnamefont{{Copeland}}}, \bibinfo{author}{\bibfnamefont{M.}~\bibnamefont{{Sami}}}, \bibnamefont{and} \bibinfo{author}{\bibfnamefont{S.}~\bibnamefont{{Tsujikawa}}}, \bibinfo{journal}{International Journal of Modern Physics D} \textbf{\bibinfo{volume}{15}}, \bibinfo{pages}{1753} (\bibinfo{year}{2006}), \eprint{hep-th/0603057}.

\bibitem[{\citenamefont{{Caldwell} and {Linder}}(2005)}]{2005quintaPhRvL..95n1301C}
\bibinfo{author}{\bibfnamefont{R.~R.} \bibnamefont{{Caldwell}}} \bibnamefont{and} \bibinfo{author}{\bibfnamefont{E.~V.} \bibnamefont{{Linder}}}, \bibinfo{journal}{Physical Review Letters} \textbf{\bibinfo{volume}{95}}, \bibinfo{eid}{141301} (\bibinfo{year}{2005}), \eprint{astro-ph/0505494}.

\bibitem[{\citenamefont{{Weinberg}}(1989)}]{1989weinbergRvMP...61....1W}
\bibinfo{author}{\bibfnamefont{S.}~\bibnamefont{{Weinberg}}}, \bibinfo{journal}{Reviews of Modern Physics} \textbf{\bibinfo{volume}{61}}, \bibinfo{pages}{1} (\bibinfo{year}{1989}).

\bibitem[{\citenamefont{Bahcall}(1999)}]{Bahcall1999}
\bibinfo{author}{\bibfnamefont{N.~A.} \bibnamefont{Bahcall}}, \bibinfo{journal}{Science} \textbf{\bibinfo{volume}{284}}, \bibinfo{pages}{1481} (\bibinfo{year}{1999}), \urlprefix\url{https://doi.org/10.1126/science.284.5419.1481}.

\bibitem[{\citenamefont{{Ostriker} and {Steinhardt}}(1995)}]{1995ostrikerNatur.377..600O}
\bibinfo{author}{\bibfnamefont{J.~P.} \bibnamefont{{Ostriker}}} \bibnamefont{and} \bibinfo{author}{\bibfnamefont{P.~J.} \bibnamefont{{Steinhardt}}}, \bibinfo{journal}{\nat} \textbf{\bibinfo{volume}{377}}, \bibinfo{pages}{600} (\bibinfo{year}{1995}).

\bibitem[{\citenamefont{Starobinsky}(1980)}]{Starobinsky1980}
\bibinfo{author}{\bibfnamefont{A.~A.} \bibnamefont{Starobinsky}}, \bibinfo{journal}{Physics Letters B} \textbf{\bibinfo{volume}{91}}, \bibinfo{pages}{99} (\bibinfo{year}{1980}), ISSN \bibinfo{issn}{03702693}.

\bibitem[{\citenamefont{Asaka et~al.}(2016)\citenamefont{Asaka, Iso, Kawai, Kohri, Noumi, and Terada}}]{staroycuerdas10.1093/ptep/ptw161}
\bibinfo{author}{\bibfnamefont{T.}~\bibnamefont{Asaka}}, \bibinfo{author}{\bibfnamefont{S.}~\bibnamefont{Iso}}, \bibinfo{author}{\bibfnamefont{H.}~\bibnamefont{Kawai}}, \bibinfo{author}{\bibfnamefont{K.}~\bibnamefont{Kohri}}, \bibinfo{author}{\bibfnamefont{T.}~\bibnamefont{Noumi}}, \bibnamefont{and} \bibinfo{author}{\bibfnamefont{T.}~\bibnamefont{Terada}}, \bibinfo{journal}{Progress of Theoretical and Experimental Physics} \textbf{\bibinfo{volume}{2016}}, \bibinfo{pages}{123E01} (\bibinfo{year}{2016}), ISSN \bibinfo{issn}{2050-3911}, \eprint{https://academic.oup.com/ptep/article-pdf/2016/12/123E01/10436310/ptw161.pdf}, \urlprefix\url{https://doi.org/10.1093/ptep/ptw161}.

\bibitem[{\citenamefont{{Cisterna} et~al.}(2024)\citenamefont{{Cisterna}, {Grandi}, and {Oliva}}}]{grandi2024PhRvD.110h4043C}
\bibinfo{author}{\bibfnamefont{A.}~\bibnamefont{{Cisterna}}}, \bibinfo{author}{\bibfnamefont{N.}~\bibnamefont{{Grandi}}}, \bibnamefont{and} \bibinfo{author}{\bibfnamefont{J.}~\bibnamefont{{Oliva}}}, \bibinfo{journal}{\prd} \textbf{\bibinfo{volume}{110}}, \bibinfo{eid}{084043} (\bibinfo{year}{2024}), \eprint{2406.10037}.

\bibitem[{\citenamefont{Nojiri and Odintsov}(2011)}]{Nojiri:2008nt}
\bibinfo{author}{\bibfnamefont{S.}~\bibnamefont{Nojiri}} \bibnamefont{and} \bibinfo{author}{\bibfnamefont{S.~D.} \bibnamefont{Odintsov}}, \bibinfo{journal}{TSPU Bulletin} \textbf{\bibinfo{volume}{N8(110)}}, \bibinfo{pages}{7} (\bibinfo{year}{2011}), \eprint{0807.0685}.

\bibitem[{\citenamefont{Artymowski and Lalak}(2014)}]{Artymowski:2014gea}
\bibinfo{author}{\bibfnamefont{M.}~\bibnamefont{Artymowski}} \bibnamefont{and} \bibinfo{author}{\bibfnamefont{Z.}~\bibnamefont{Lalak}}, \bibinfo{journal}{JCAP} \textbf{\bibinfo{volume}{09}}, \bibinfo{pages}{036} (\bibinfo{year}{2014}), \eprint{1405.7818}.

\bibitem[{\citenamefont{Hawking et~al.}(2001)\citenamefont{Hawking, Hertog, and Reall}}]{hawkingPhysRevD.63.083504}
\bibinfo{author}{\bibfnamefont{S.~W.} \bibnamefont{Hawking}}, \bibinfo{author}{\bibfnamefont{T.}~\bibnamefont{Hertog}}, \bibnamefont{and} \bibinfo{author}{\bibfnamefont{H.~S.} \bibnamefont{Reall}}, \bibinfo{journal}{Phys. Rev. D} \textbf{\bibinfo{volume}{63}}, \bibinfo{pages}{083504} (\bibinfo{year}{2001}), \urlprefix\url{https://link.aps.org/doi/10.1103/PhysRevD.63.083504}.

\bibitem[{\citenamefont{{De Felice} and {Tsujikawa}}(2010{\natexlab{a}})}]{defeliytsuji2010LRR....13....3D}
\bibinfo{author}{\bibfnamefont{A.}~\bibnamefont{{De Felice}}} \bibnamefont{and} \bibinfo{author}{\bibfnamefont{S.}~\bibnamefont{{Tsujikawa}}}, \bibinfo{journal}{Living Reviews in Relativity} \textbf{\bibinfo{volume}{13}}, \bibinfo{eid}{3} (\bibinfo{year}{2010}{\natexlab{a}}), \eprint{1002.4928}.

\bibitem[{\citenamefont{D.{Perez}}(2023)}]{2023danielaFdeTtgif.book.....A}
\bibinfo{author}{\bibfnamefont{.}~\bibnamefont{D.{Perez}}, \bibfnamefont{et~al.}}, \emph{\bibinfo{title}{{New Phenomena and New States of Matter in the Universe}}} (\bibinfo{year}{2023}).

\bibitem[{\citenamefont{{Dicke}}(1962)}]{1962dickePhRv..125.2163D}
\bibinfo{author}{\bibfnamefont{R.~H.} \bibnamefont{{Dicke}}}, \bibinfo{journal}{Physical Review} \textbf{\bibinfo{volume}{125}}, \bibinfo{pages}{2163} (\bibinfo{year}{1962}).

\bibitem[{\citenamefont{{Capozziello} et~al.}(1997)\citenamefont{{Capozziello}, {de Ritis}, and {Marino}}}]{capozziello1997CQGra..14.3243C}
\bibinfo{author}{\bibfnamefont{S.}~\bibnamefont{{Capozziello}}}, \bibinfo{author}{\bibfnamefont{R.}~\bibnamefont{{de Ritis}}}, \bibnamefont{and} \bibinfo{author}{\bibfnamefont{A.~A.} \bibnamefont{{Marino}}}, \bibinfo{journal}{Classical and Quantum Gravity} \textbf{\bibinfo{volume}{14}}, \bibinfo{pages}{3243} (\bibinfo{year}{1997}), \eprint{gr-qc/9612053}.

\bibitem[{\citenamefont{{Capozziello} et~al.}(2010)\citenamefont{{Capozziello}, {Martin-Moruno}, and {Rubano}}}]{capozziello2010PhLB..689..117C}
\bibinfo{author}{\bibfnamefont{S.}~\bibnamefont{{Capozziello}}}, \bibinfo{author}{\bibfnamefont{P.}~\bibnamefont{{Martin-Moruno}}}, \bibnamefont{and} \bibinfo{author}{\bibfnamefont{C.}~\bibnamefont{{Rubano}}}, \bibinfo{journal}{Physics Letters B} \textbf{\bibinfo{volume}{689}}, \bibinfo{pages}{117} (\bibinfo{year}{2010}), \eprint{1003.5394}.

\bibitem[{\citenamefont{Park and Lee}(2023)}]{Park:2023aht}
\bibinfo{author}{\bibfnamefont{J.}~\bibnamefont{Park}} \bibnamefont{and} \bibinfo{author}{\bibfnamefont{T.~H.} \bibnamefont{Lee}}, \bibinfo{journal}{Mod. Phys. Lett. A} \textbf{\bibinfo{volume}{38}}, \bibinfo{pages}{2350047} (\bibinfo{year}{2023}).

\bibitem[{\citenamefont{{Hu} et~al.}(2016)\citenamefont{{Hu}, {Raveri}, {Rizzato}, and {Silvestri}}}]{2016MNRAS.459.3880H}
\bibinfo{author}{\bibfnamefont{B.}~\bibnamefont{{Hu}}}, \bibinfo{author}{\bibfnamefont{M.}~\bibnamefont{{Raveri}}}, \bibinfo{author}{\bibfnamefont{M.}~\bibnamefont{{Rizzato}}}, \bibnamefont{and} \bibinfo{author}{\bibfnamefont{A.}~\bibnamefont{{Silvestri}}}, \bibinfo{journal}{\mnras} \textbf{\bibinfo{volume}{459}}, \bibinfo{pages}{3880} (\bibinfo{year}{2016}), \eprint{1601.07536}.

\bibitem[{\citenamefont{Nunes~et al.}(2017)}]{Nunes_2017}
\bibinfo{author}{\bibfnamefont{.}~\bibnamefont{Nunes~et al.}}, \bibinfo{journal}{Journal of Cosmology and Astroparticle Physics} \textbf{\bibinfo{volume}{2017}}, \bibinfo{pages}{005} (\bibinfo{year}{2017}), \urlprefix\url{https://doi.org/10.1088%2F1475-7516%2F2017%2F01%2F005}.

\bibitem[{\citenamefont{{Odintsov} et~al.}(2017)\citenamefont{{Odintsov}, {S{\'a}ez-Chill{\'o}n G{\'o}mez}, and {Sharov}}}]{Odintsov2017}
\bibinfo{author}{\bibfnamefont{S.~D.} \bibnamefont{{Odintsov}}}, \bibinfo{author}{\bibfnamefont{D.}~\bibnamefont{{S{\'a}ez-Chill{\'o}n G{\'o}mez}}}, \bibnamefont{and} \bibinfo{author}{\bibfnamefont{G.~S.} \bibnamefont{{Sharov}}}, \bibinfo{journal}{Eur. Phys. J. C} \textbf{\bibinfo{volume}{77}}, \bibinfo{eid}{862} (\bibinfo{year}{2017}).

\bibitem[{\citenamefont{{de la Cruz-Dombriz} et~al.}(2016)\citenamefont{{de la Cruz-Dombriz}, {Dunsby}, {Kandhai}, and {S{\'a}ez-G{\'o}mez}}}]{2016PhRvD..93h4016D}
\bibinfo{author}{\bibfnamefont{{\'A}.}~\bibnamefont{{de la Cruz-Dombriz}}}, \bibinfo{author}{\bibfnamefont{P.~K.~S.} \bibnamefont{{Dunsby}}}, \bibinfo{author}{\bibfnamefont{S.}~\bibnamefont{{Kandhai}}}, \bibnamefont{and} \bibinfo{author}{\bibfnamefont{D.}~\bibnamefont{{S{\'a}ez-G{\'o}mez}}}, \bibinfo{journal}{\prd} \textbf{\bibinfo{volume}{93}}, \bibinfo{eid}{084016} (\bibinfo{year}{2016}), \eprint{1511.00102}.

\bibitem[{\citenamefont{{Leizerovich} et~al.}(2022)\citenamefont{{Leizerovich}, {Kraiselburd}, {Landau}, and {Sc{\'o}ccola}}}]{M2022PhRvD.105j3526L}
\bibinfo{author}{\bibfnamefont{M.}~\bibnamefont{{Leizerovich}}}, \bibinfo{author}{\bibfnamefont{L.}~\bibnamefont{{Kraiselburd}}}, \bibinfo{author}{\bibfnamefont{S.}~\bibnamefont{{Landau}}}, \bibnamefont{and} \bibinfo{author}{\bibfnamefont{C.~G.} \bibnamefont{{Sc{\'o}ccola}}}, \bibinfo{journal}{Physical Review D} \textbf{\bibinfo{volume}{105}}, \bibinfo{eid}{103526} (\bibinfo{year}{2022}), \eprint{2112.01492}.

\bibitem[{\citenamefont{Odintsov et~al.}(2021)\citenamefont{Odintsov, {Sáez-Chillón Gómez}, and Sharov}}]{ODINTSOV2021115377}
\bibinfo{author}{\bibfnamefont{S.~D.} \bibnamefont{Odintsov}}, \bibinfo{author}{\bibfnamefont{D.}~\bibnamefont{{Sáez-Chillón Gómez}}}, \bibnamefont{and} \bibinfo{author}{\bibfnamefont{G.~S.} \bibnamefont{Sharov}}, \bibinfo{journal}{Nuclear Physics B} \textbf{\bibinfo{volume}{966}}, \bibinfo{pages}{115377} (\bibinfo{year}{2021}), ISSN \bibinfo{issn}{0550-3213}, \urlprefix\url{https://www.sciencedirect.com/science/article/pii/S0550321321000742}.

\bibitem[{\citenamefont{{Farrugia} et~al.}(2021)\citenamefont{{Farrugia}, {Sultana}, and {Mifsud}}}]{furrugia2021PhRvD.104l3503F}
\bibinfo{author}{\bibfnamefont{C.~R.} \bibnamefont{{Farrugia}}}, \bibinfo{author}{\bibfnamefont{J.}~\bibnamefont{{Sultana}}}, \bibnamefont{and} \bibinfo{author}{\bibfnamefont{J.}~\bibnamefont{{Mifsud}}}, \bibinfo{journal}{\prd} \textbf{\bibinfo{volume}{104}}, \bibinfo{eid}{123503} (\bibinfo{year}{2021}), \eprint{2106.04657}.

\bibitem[{\citenamefont{{Ravi} et~al.}(2024{\natexlab{a}})\citenamefont{{Ravi}, {Chatterjee}, {Jana}, and {Bandyopadhyay}}}]{2024MNRAS.527.7626R}
\bibinfo{author}{\bibfnamefont{K.}~\bibnamefont{{Ravi}}}, \bibinfo{author}{\bibfnamefont{A.}~\bibnamefont{{Chatterjee}}}, \bibinfo{author}{\bibfnamefont{B.}~\bibnamefont{{Jana}}}, \bibnamefont{and} \bibinfo{author}{\bibfnamefont{A.}~\bibnamefont{{Bandyopadhyay}}}, \bibinfo{journal}{Monthly Notices of the Royal Astronomical Society} \textbf{\bibinfo{volume}{527}}, \bibinfo{pages}{7626} (\bibinfo{year}{2024}{\natexlab{a}}), \eprint{2306.12585}.

\bibitem[{\citenamefont{{Kou} et~al.}(2024)\citenamefont{{Kou}, {Murray}, and {Bartlett}}}]{2024A&A...686A.193K}
\bibinfo{author}{\bibfnamefont{R.}~\bibnamefont{{Kou}}}, \bibinfo{author}{\bibfnamefont{C.}~\bibnamefont{{Murray}}}, \bibnamefont{and} \bibinfo{author}{\bibfnamefont{J.~G.} \bibnamefont{{Bartlett}}}, \bibinfo{journal}{\aap} \textbf{\bibinfo{volume}{686}}, \bibinfo{eid}{A193} (\bibinfo{year}{2024}), \eprint{2311.09936}.

\bibitem[{\citenamefont{{Li} et~al.}(2025)\citenamefont{{Li}, {Riess}, {Scolnic}, {Casertano}, and {Anand}}}]{2025arXiv250205259L}
\bibinfo{author}{\bibfnamefont{S.}~\bibnamefont{{Li}}}, \bibinfo{author}{\bibfnamefont{A.~G.} \bibnamefont{{Riess}}}, \bibinfo{author}{\bibfnamefont{D.}~\bibnamefont{{Scolnic}}}, \bibinfo{author}{\bibfnamefont{S.}~\bibnamefont{{Casertano}}}, \bibnamefont{and} \bibinfo{author}{\bibfnamefont{G.~S.} \bibnamefont{{Anand}}}, \bibinfo{journal}{arXiv e-prints} \bibinfo{eid}{arXiv:2502.05259} (\bibinfo{year}{2025}), \eprint{2502.05259}.

\bibitem[{\citenamefont{{Plaza} and {Kraiselburd}}(2025)}]{plaza2025arXiv250316132P}
\bibinfo{author}{\bibfnamefont{F.}~\bibnamefont{{Plaza}}} \bibnamefont{and} \bibinfo{author}{\bibfnamefont{L.}~\bibnamefont{{Kraiselburd}}}, \bibinfo{journal}{arXiv e-prints} \bibinfo{eid}{arXiv:2503.16132} (\bibinfo{year}{2025}), \eprint{2503.16132}.

\bibitem[{\citenamefont{{Ravi} et~al.}(2024{\natexlab{b}})\citenamefont{{Ravi}, {Chatterjee}, {Jana}, and {Bandyopadhyay}}}]{ravi2024MNRAS.527.7626R}
\bibinfo{author}{\bibfnamefont{K.}~\bibnamefont{{Ravi}}}, \bibinfo{author}{\bibfnamefont{A.}~\bibnamefont{{Chatterjee}}}, \bibinfo{author}{\bibfnamefont{B.}~\bibnamefont{{Jana}}}, \bibnamefont{and} \bibinfo{author}{\bibfnamefont{A.}~\bibnamefont{{Bandyopadhyay}}}, \bibinfo{journal}{\mnras} \textbf{\bibinfo{volume}{527}}, \bibinfo{pages}{7626} (\bibinfo{year}{2024}{\natexlab{b}}), \eprint{2306.12585}.

\bibitem[{\citenamefont{Chantada et~al.}(2024)\citenamefont{Chantada, Landau, Protopapas, Sc\'occola, and Garraffo}}]{PhysRevD.109.123514}
\bibinfo{author}{\bibfnamefont{A.~T.} \bibnamefont{Chantada}}, \bibinfo{author}{\bibfnamefont{S.~J.} \bibnamefont{Landau}}, \bibinfo{author}{\bibfnamefont{P.}~\bibnamefont{Protopapas}}, \bibinfo{author}{\bibfnamefont{C.~G.} \bibnamefont{Sc\'occola}}, \bibnamefont{and} \bibinfo{author}{\bibfnamefont{C.}~\bibnamefont{Garraffo}}, \bibinfo{journal}{Phys. Rev. D} \textbf{\bibinfo{volume}{109}}, \bibinfo{pages}{123514} (\bibinfo{year}{2024}), \urlprefix\url{https://link.aps.org/doi/10.1103/PhysRevD.109.123514}.

\bibitem[{\citenamefont{{DESI Collaboration} et~al.}(2025)\citenamefont{{DESI Collaboration}, {Abdul Karim}, {Aguilar}, {Ahlen}, {Alam}, {Allen}, {Allende Prieto}, {Alves}, {Anand}, {Andrade} et~al.}}]{DR22025arXiv250314738D}
\bibinfo{author}{\bibnamefont{{DESI Collaboration}}}, \bibinfo{author}{\bibfnamefont{M.}~\bibnamefont{{Abdul Karim}}}, \bibinfo{author}{\bibfnamefont{J.}~\bibnamefont{{Aguilar}}}, \bibinfo{author}{\bibfnamefont{S.}~\bibnamefont{{Ahlen}}}, \bibinfo{author}{\bibfnamefont{S.}~\bibnamefont{{Alam}}}, \bibinfo{author}{\bibfnamefont{L.}~\bibnamefont{{Allen}}}, \bibinfo{author}{\bibfnamefont{C.}~\bibnamefont{{Allende Prieto}}}, \bibinfo{author}{\bibfnamefont{O.}~\bibnamefont{{Alves}}}, \bibinfo{author}{\bibfnamefont{A.}~\bibnamefont{{Anand}}}, \bibinfo{author}{\bibfnamefont{U.}~\bibnamefont{{Andrade}}}, \bibnamefont{et~al.}, \bibinfo{journal}{arXiv e-prints} \bibinfo{eid}{arXiv:2503.14738} (\bibinfo{year}{2025}), \eprint{2503.14738}.

\bibitem[{\citenamefont{{Amendola} and {Tsujikawa}}(2010)}]{cond12010deto.book.....A}
\bibinfo{author}{\bibfnamefont{L.}~\bibnamefont{{Amendola}}} \bibnamefont{and} \bibinfo{author}{\bibfnamefont{S.}~\bibnamefont{{Tsujikawa}}}, \emph{\bibinfo{title}{{Dark Energy: Theory and Observations}}} (\bibinfo{year}{2010}).

\bibitem[{\citenamefont{{De Felice} and {Tsujikawa}}(2010{\natexlab{b}})}]{cond22010LRR....13....3D}
\bibinfo{author}{\bibfnamefont{A.}~\bibnamefont{{De Felice}}} \bibnamefont{and} \bibinfo{author}{\bibfnamefont{S.}~\bibnamefont{{Tsujikawa}}}, \bibinfo{journal}{Living Reviews in Relativity} \textbf{\bibinfo{volume}{13}}, \bibinfo{eid}{3} (\bibinfo{year}{2010}{\natexlab{b}}), \eprint{1002.4928}.

\bibitem[{\citenamefont{{Hu} and {Sawicki}}(2007)}]{husawicki2007PhRvD..76f4004H}
\bibinfo{author}{\bibfnamefont{W.}~\bibnamefont{{Hu}}} \bibnamefont{and} \bibinfo{author}{\bibfnamefont{I.}~\bibnamefont{{Sawicki}}}, \bibinfo{journal}{\prd} \textbf{\bibinfo{volume}{76}}, \bibinfo{eid}{064004} (\bibinfo{year}{2007}), \eprint{0705.1158}.

\bibitem[{\citenamefont{Cognola et~al.}(2008)\citenamefont{Cognola, Elizalde, Nojiri, Odintsov, Sebastiani, and Zerbini}}]{expPhysRevD.77.046009}
\bibinfo{author}{\bibfnamefont{G.}~\bibnamefont{Cognola}}, \bibinfo{author}{\bibfnamefont{E.}~\bibnamefont{Elizalde}}, \bibinfo{author}{\bibfnamefont{S.}~\bibnamefont{Nojiri}}, \bibinfo{author}{\bibfnamefont{S.~D.} \bibnamefont{Odintsov}}, \bibinfo{author}{\bibfnamefont{L.}~\bibnamefont{Sebastiani}}, \bibnamefont{and} \bibinfo{author}{\bibfnamefont{S.}~\bibnamefont{Zerbini}}, \bibinfo{journal}{Phys. Rev. D} \textbf{\bibinfo{volume}{77}}, \bibinfo{pages}{046009} (\bibinfo{year}{2008}), \urlprefix\url{https://link.aps.org/doi/10.1103/PhysRevD.77.046009}.

\bibitem[{\citenamefont{Tsujikawa}(2008)}]{Tsujikawa_2008}
\bibinfo{author}{\bibfnamefont{S.}~\bibnamefont{Tsujikawa}}, \bibinfo{journal}{Physical Review D} \textbf{\bibinfo{volume}{77}} (\bibinfo{year}{2008}), ISSN \bibinfo{issn}{1550-2368}, \urlprefix\url{http://dx.doi.org/10.1103/PhysRevD.77.023507}.

\bibitem[{\citenamefont{{Simon} et~al.}(2005)\citenamefont{{Simon}, {Verde}, and {Jimenez}}}]{simon05}
\bibinfo{author}{\bibfnamefont{J.}~\bibnamefont{{Simon}}}, \bibinfo{author}{\bibfnamefont{L.}~\bibnamefont{{Verde}}}, \bibnamefont{and} \bibinfo{author}{\bibfnamefont{R.}~\bibnamefont{{Jimenez}}}, \bibinfo{journal}{\prd} \textbf{\bibinfo{volume}{71}}, \bibinfo{eid}{123001} (\bibinfo{year}{2005}), \eprint{astro-ph/0412269}.

\bibitem[{\citenamefont{{Stern} et~al.}(2010)\citenamefont{{Stern}, {Jimenez}, {Verde}, {Kamionkowski}, and {Stanford}}}]{stern10}
\bibinfo{author}{\bibfnamefont{D.}~\bibnamefont{{Stern}}}, \bibinfo{author}{\bibfnamefont{R.}~\bibnamefont{{Jimenez}}}, \bibinfo{author}{\bibfnamefont{L.}~\bibnamefont{{Verde}}}, \bibinfo{author}{\bibfnamefont{M.}~\bibnamefont{{Kamionkowski}}}, \bibnamefont{and} \bibinfo{author}{\bibfnamefont{S.~A.} \bibnamefont{{Stanford}}}, \bibinfo{journal}{\jcap} \textbf{\bibinfo{volume}{2}}, \bibinfo{eid}{008} (\bibinfo{year}{2010}), \eprint{0907.3149}.

\bibitem[{\citenamefont{{Moresco} et~al.}(2012)\citenamefont{{Moresco}, {Cimatti}, {Jimenez}, {Pozzetti}, {Zamorani}, {Bolzonella}, {Dunlop}, {Lamareille}, {Mignoli}, {Pearce} et~al.}}]{moresco12}
\bibinfo{author}{\bibfnamefont{M.}~\bibnamefont{{Moresco}}}, \bibinfo{author}{\bibfnamefont{A.}~\bibnamefont{{Cimatti}}}, \bibinfo{author}{\bibfnamefont{R.}~\bibnamefont{{Jimenez}}}, \bibinfo{author}{\bibfnamefont{L.}~\bibnamefont{{Pozzetti}}}, \bibinfo{author}{\bibfnamefont{G.}~\bibnamefont{{Zamorani}}}, \bibinfo{author}{\bibfnamefont{M.}~\bibnamefont{{Bolzonella}}}, \bibinfo{author}{\bibfnamefont{J.}~\bibnamefont{{Dunlop}}}, \bibinfo{author}{\bibfnamefont{F.}~\bibnamefont{{Lamareille}}}, \bibinfo{author}{\bibfnamefont{M.}~\bibnamefont{{Mignoli}}}, \bibinfo{author}{\bibfnamefont{H.}~\bibnamefont{{Pearce}}}, \bibnamefont{et~al.}, \bibinfo{journal}{\jcap} \textbf{\bibinfo{volume}{8}}, \bibinfo{eid}{006} (\bibinfo{year}{2012}), \eprint{1201.3609}.

\bibitem[{\citenamefont{{Zhang} et~al.}(2014)\citenamefont{{Zhang}, {Zhang}, {Yuan}, {Liu}, {Zhang}, and {Sun}}}]{zhang14}
\bibinfo{author}{\bibfnamefont{C.}~\bibnamefont{{Zhang}}}, \bibinfo{author}{\bibfnamefont{H.}~\bibnamefont{{Zhang}}}, \bibinfo{author}{\bibfnamefont{S.}~\bibnamefont{{Yuan}}}, \bibinfo{author}{\bibfnamefont{S.}~\bibnamefont{{Liu}}}, \bibinfo{author}{\bibfnamefont{T.-J.} \bibnamefont{{Zhang}}}, \bibnamefont{and} \bibinfo{author}{\bibfnamefont{Y.-C.} \bibnamefont{{Sun}}}, \bibinfo{journal}{Research in Astronomy and Astrophysics} \textbf{\bibinfo{volume}{14}}, \bibinfo{eid}{1221-1233} (\bibinfo{year}{2014}), \eprint{1207.4541}.

\bibitem[{\citenamefont{{Moresco}}(2015)}]{moresco15}
\bibinfo{author}{\bibfnamefont{M.}~\bibnamefont{{Moresco}}}, \bibinfo{journal}{\mnras} \textbf{\bibinfo{volume}{450}}, \bibinfo{pages}{L16} (\bibinfo{year}{2015}), \eprint{1503.01116}.

\bibitem[{\citenamefont{{Moresco} et~al.}(2016)\citenamefont{{Moresco}, {Pozzetti}, {Cimatti}, {Jimenez}, {Maraston}, {Verde}, {Thomas}, {Citro}, {Tojeiro}, and {Wilkinson}}}]{CC2}
\bibinfo{author}{\bibfnamefont{M.}~\bibnamefont{{Moresco}}}, \bibinfo{author}{\bibfnamefont{L.}~\bibnamefont{{Pozzetti}}}, \bibinfo{author}{\bibfnamefont{A.}~\bibnamefont{{Cimatti}}}, \bibinfo{author}{\bibfnamefont{R.}~\bibnamefont{{Jimenez}}}, \bibinfo{author}{\bibfnamefont{C.}~\bibnamefont{{Maraston}}}, \bibinfo{author}{\bibfnamefont{L.}~\bibnamefont{{Verde}}}, \bibinfo{author}{\bibfnamefont{D.}~\bibnamefont{{Thomas}}}, \bibinfo{author}{\bibfnamefont{A.}~\bibnamefont{{Citro}}}, \bibinfo{author}{\bibfnamefont{R.}~\bibnamefont{{Tojeiro}}}, \bibnamefont{and} \bibinfo{author}{\bibfnamefont{D.}~\bibnamefont{{Wilkinson}}}, \bibinfo{journal}{\jcap} \textbf{\bibinfo{volume}{5}}, \bibinfo{eid}{014} (\bibinfo{year}{2016}), \eprint{1601.01701}.

\bibitem[{\citenamefont{Borghi et~al.}(2022)\citenamefont{Borghi, Moresco, and Cimatti}}]{Borghi_2022}
\bibinfo{author}{\bibfnamefont{N.}~\bibnamefont{Borghi}}, \bibinfo{author}{\bibfnamefont{M.}~\bibnamefont{Moresco}}, \bibnamefont{and} \bibinfo{author}{\bibfnamefont{A.}~\bibnamefont{Cimatti}}, \bibinfo{journal}{The Astrophysical Journal Letters} \textbf{\bibinfo{volume}{928}}, \bibinfo{pages}{L4} (\bibinfo{year}{2022}), \urlprefix\url{https://dx.doi.org/10.3847/2041-8213/ac3fb2}.

\bibitem[{\citenamefont{{Brout} et~al.}(2022)\citenamefont{{Brout}, {Scolnic}, and et~al.}}]{2022ApJ...938..110B}
\bibinfo{author}{\bibfnamefont{D.}~\bibnamefont{{Brout}}}, \bibinfo{author}{\bibfnamefont{D.}~\bibnamefont{{Scolnic}}}, \bibnamefont{and} \bibinfo{author}{\bibnamefont{et~al.}}, \bibinfo{journal}{The Astrophys. Journal} \textbf{\bibinfo{volume}{938}}, \bibinfo{eid}{110} (\bibinfo{year}{2022}), \eprint{2202.04077}.

\bibitem[{\citenamefont{{Scolnic} et~al.}(2022)\citenamefont{{Scolnic}, {Brout}, {Carr}, and et~al.}}]{2022ApJ...938..113S}
\bibinfo{author}{\bibfnamefont{D.}~\bibnamefont{{Scolnic}}}, \bibinfo{author}{\bibfnamefont{D.}~\bibnamefont{{Brout}}}, \bibinfo{author}{\bibfnamefont{A.}~\bibnamefont{{Carr}}}, \bibnamefont{and} \bibinfo{author}{\bibnamefont{et~al.}}, \bibinfo{journal}{The Astrophys. Journal} \textbf{\bibinfo{volume}{938}}, \bibinfo{eid}{113} (\bibinfo{year}{2022}), \eprint{2112.03863}.

\bibitem[{\citenamefont{{Riess} et~al.}(2022)\citenamefont{{Riess}, {Yuan}, {Macri}, {Scolnic}, {Brout}, {Casertano}, {Jones}, {Murakami}, {Anand}, {Breuval} et~al.}}]{riess2022ApJ...934L...7R}
\bibinfo{author}{\bibfnamefont{A.~G.} \bibnamefont{{Riess}}}, \bibinfo{author}{\bibfnamefont{W.}~\bibnamefont{{Yuan}}}, \bibinfo{author}{\bibfnamefont{L.~M.} \bibnamefont{{Macri}}}, \bibinfo{author}{\bibfnamefont{D.}~\bibnamefont{{Scolnic}}}, \bibinfo{author}{\bibfnamefont{D.}~\bibnamefont{{Brout}}}, \bibinfo{author}{\bibfnamefont{S.}~\bibnamefont{{Casertano}}}, \bibinfo{author}{\bibfnamefont{D.~O.} \bibnamefont{{Jones}}}, \bibinfo{author}{\bibfnamefont{Y.}~\bibnamefont{{Murakami}}}, \bibinfo{author}{\bibfnamefont{G.~S.} \bibnamefont{{Anand}}}, \bibinfo{author}{\bibfnamefont{L.}~\bibnamefont{{Breuval}}}, \bibnamefont{et~al.}, \bibinfo{journal}{\apjl} \textbf{\bibinfo{volume}{934}}, \bibinfo{eid}{L7} (\bibinfo{year}{2022}), \eprint{2112.04510}.

\bibitem[{\citenamefont{{Conley} et~al.}(2011)\citenamefont{{Conley}, {Guy}, {Sullivan}, {Regnault}, {Astier}, {Balland}, {Basa}, {Carlberg}, {Fouchez}, {Hardin} et~al.}}]{Conley}
\bibinfo{author}{\bibfnamefont{A.}~\bibnamefont{{Conley}}}, \bibinfo{author}{\bibfnamefont{J.}~\bibnamefont{{Guy}}}, \bibinfo{author}{\bibfnamefont{M.}~\bibnamefont{{Sullivan}}}, \bibinfo{author}{\bibfnamefont{N.}~\bibnamefont{{Regnault}}}, \bibinfo{author}{\bibfnamefont{P.}~\bibnamefont{{Astier}}}, \bibinfo{author}{\bibfnamefont{C.}~\bibnamefont{{Balland}}}, \bibinfo{author}{\bibfnamefont{S.}~\bibnamefont{{Basa}}}, \bibinfo{author}{\bibfnamefont{R.~G.} \bibnamefont{{Carlberg}}}, \bibinfo{author}{\bibfnamefont{D.}~\bibnamefont{{Fouchez}}}, \bibinfo{author}{\bibfnamefont{D.}~\bibnamefont{{Hardin}}}, \bibnamefont{et~al.}, \bibinfo{journal}{Astropys. Journal Suppl.} \textbf{\bibinfo{volume}{192}}, \bibinfo{eid}{1} (\bibinfo{year}{2011}), \eprint{1104.1443}.

\bibitem[{\citenamefont{{Camarena} and {Marra}}(2020)}]{Camarena2020}
\bibinfo{author}{\bibfnamefont{D.}~\bibnamefont{{Camarena}}} \bibnamefont{and} \bibinfo{author}{\bibfnamefont{V.}~\bibnamefont{{Marra}}}, \bibinfo{journal}{Physical Review Research} \textbf{\bibinfo{volume}{2}}, \bibinfo{eid}{013028} (\bibinfo{year}{2020}), \eprint{1906.11814}.

\bibitem[{\citenamefont{{Camarena} and {Marra}}(2023)}]{Camarena2023}
\bibinfo{author}{\bibfnamefont{D.}~\bibnamefont{{Camarena}}} \bibnamefont{and} \bibinfo{author}{\bibfnamefont{V.}~\bibnamefont{{Marra}}}, \bibinfo{journal}{arXiv e-prints} \bibinfo{eid}{arXiv:2307.02434} (\bibinfo{year}{2023}), \eprint{2307.02434}.

\bibitem[{\citenamefont{{Chantada} et~al.}(2023)\citenamefont{{Chantada}, {Landau}, {Protopapas}, {Sc{\'o}ccola}, and {Garraffo}}}]{2023PhRvD.107f3523C}
\bibinfo{author}{\bibfnamefont{A.~T.} \bibnamefont{{Chantada}}}, \bibinfo{author}{\bibfnamefont{S.~J.} \bibnamefont{{Landau}}}, \bibinfo{author}{\bibfnamefont{P.}~\bibnamefont{{Protopapas}}}, \bibinfo{author}{\bibfnamefont{C.~G.} \bibnamefont{{Sc{\'o}ccola}}}, \bibnamefont{and} \bibinfo{author}{\bibfnamefont{C.}~\bibnamefont{{Garraffo}}}, \bibinfo{journal}{Phys. Rev. D} \textbf{\bibinfo{volume}{107}}, \bibinfo{eid}{063523} (\bibinfo{year}{2023}), \eprint{2205.02945}.

\bibitem[{\citenamefont{{Freedman} et~al.}(2024)\citenamefont{{Freedman}, {Madore}, {Jang}, {Hoyt}, {Lee}, and {Owens}}}]{2024arXiv240806153F}
\bibinfo{author}{\bibfnamefont{W.~L.} \bibnamefont{{Freedman}}}, \bibinfo{author}{\bibfnamefont{B.~F.} \bibnamefont{{Madore}}}, \bibinfo{author}{\bibfnamefont{I.~S.} \bibnamefont{{Jang}}}, \bibinfo{author}{\bibfnamefont{T.~J.} \bibnamefont{{Hoyt}}}, \bibinfo{author}{\bibfnamefont{A.~J.} \bibnamefont{{Lee}}}, \bibnamefont{and} \bibinfo{author}{\bibfnamefont{K.~A.} \bibnamefont{{Owens}}}, \bibinfo{journal}{arXiv e-prints} \bibinfo{eid}{arXiv:2408.06153} (\bibinfo{year}{2024}), \eprint{2408.06153}.

\bibitem[{\citenamefont{{Perivolaropoulos}}(2024)}]{peri2024PhRvD.110l3518P}
\bibinfo{author}{\bibfnamefont{L.}~\bibnamefont{{Perivolaropoulos}}}, \bibinfo{journal}{\prd} \textbf{\bibinfo{volume}{110}}, \bibinfo{eid}{123518} (\bibinfo{year}{2024}), \eprint{2408.11031}.

\bibitem[{\citenamefont{Ross et~al.}(2015)\citenamefont{Ross, Samushia, Howlett, Percival, Burden, and Manera}}]{SDSS_DR7}
\bibinfo{author}{\bibfnamefont{A.~J.} \bibnamefont{Ross}}, \bibinfo{author}{\bibfnamefont{L.}~\bibnamefont{Samushia}}, \bibinfo{author}{\bibfnamefont{C.}~\bibnamefont{Howlett}}, \bibinfo{author}{\bibfnamefont{W.~J.} \bibnamefont{Percival}}, \bibinfo{author}{\bibfnamefont{A.}~\bibnamefont{Burden}}, \bibnamefont{and} \bibinfo{author}{\bibfnamefont{M.}~\bibnamefont{Manera}}, \bibinfo{journal}{Monthly Notices of the Royal Astronomical Society} \textbf{\bibinfo{volume}{449}}, \bibinfo{pages}{835–847} (\bibinfo{year}{2015}), ISSN \bibinfo{issn}{0035-8711}, \urlprefix\url{http://dx.doi.org/10.1093/mnras/stv154}.

\bibitem[{\citenamefont{Bautista et~al.}(2017)\citenamefont{Bautista, Busca, Guy, Rich, Blomqvist, du~Mas~des Bourboux, Pieri, Font-Ribera, Bailey, Delubac et~al.}}]{SDSSIIILaforests}
\bibinfo{author}{\bibfnamefont{J.~E.} \bibnamefont{Bautista}}, \bibinfo{author}{\bibfnamefont{N.~G.} \bibnamefont{Busca}}, \bibinfo{author}{\bibfnamefont{J.}~\bibnamefont{Guy}}, \bibinfo{author}{\bibfnamefont{J.}~\bibnamefont{Rich}}, \bibinfo{author}{\bibfnamefont{M.}~\bibnamefont{Blomqvist}}, \bibinfo{author}{\bibfnamefont{H.}~\bibnamefont{du~Mas~des Bourboux}}, \bibinfo{author}{\bibfnamefont{M.~M.} \bibnamefont{Pieri}}, \bibinfo{author}{\bibfnamefont{A.}~\bibnamefont{Font-Ribera}}, \bibinfo{author}{\bibfnamefont{S.}~\bibnamefont{Bailey}}, \bibinfo{author}{\bibfnamefont{T.}~\bibnamefont{Delubac}}, \bibnamefont{et~al.}, \bibinfo{journal}{Astronomy and Astrophysics} \textbf{\bibinfo{volume}{603}}, \bibinfo{pages}{A12} (\bibinfo{year}{2017}), ISSN \bibinfo{issn}{1432-0746}, \urlprefix\url{http://dx.doi.org/10.1051/0004-6361/201730533}.

\bibitem[{\citenamefont{Bautista et~al.}(2020)\citenamefont{Bautista, Paviot, Vargas~Magaña, de~la Torre, Fromenteau, Gil-Marín, Ross, Burtin, Dawson, Hou et~al.}}]{SDSSIVLRGanisotropicCF}
\bibinfo{author}{\bibfnamefont{J.~E.} \bibnamefont{Bautista}}, \bibinfo{author}{\bibfnamefont{R.}~\bibnamefont{Paviot}}, \bibinfo{author}{\bibfnamefont{M.}~\bibnamefont{Vargas~Magaña}}, \bibinfo{author}{\bibfnamefont{S.}~\bibnamefont{de~la Torre}}, \bibinfo{author}{\bibfnamefont{S.}~\bibnamefont{Fromenteau}}, \bibinfo{author}{\bibfnamefont{H.}~\bibnamefont{Gil-Marín}}, \bibinfo{author}{\bibfnamefont{A.~J.} \bibnamefont{Ross}}, \bibinfo{author}{\bibfnamefont{E.}~\bibnamefont{Burtin}}, \bibinfo{author}{\bibfnamefont{K.~S.} \bibnamefont{Dawson}}, \bibinfo{author}{\bibfnamefont{J.}~\bibnamefont{Hou}}, \bibnamefont{et~al.}, \bibinfo{journal}{Monthly Notices of the Royal Astronomical Society} \textbf{\bibinfo{volume}{500}}, \bibinfo{pages}{736–762} (\bibinfo{year}{2020}), ISSN \bibinfo{issn}{1365-2966}, \urlprefix\url{http://dx.doi.org/10.1093/mnras/staa2800}.

\bibitem[{\citenamefont{{Neveux} et~al.}(2020)\citenamefont{{Neveux}, {Burtin}, {de Mattia}, {Smith}, {Ross}, {Hou}, {Bautista}, {Brinkmann}, {Chuang}, {Dawson} et~al.}}]{SDSSIVQSOanisotropicPS}
\bibinfo{author}{\bibfnamefont{R.}~\bibnamefont{{Neveux}}}, \bibinfo{author}{\bibfnamefont{E.}~\bibnamefont{{Burtin}}}, \bibinfo{author}{\bibfnamefont{A.}~\bibnamefont{{de Mattia}}}, \bibinfo{author}{\bibfnamefont{A.}~\bibnamefont{{Smith}}}, \bibinfo{author}{\bibfnamefont{A.~J.} \bibnamefont{{Ross}}}, \bibinfo{author}{\bibfnamefont{J.}~\bibnamefont{{Hou}}}, \bibinfo{author}{\bibfnamefont{J.}~\bibnamefont{{Bautista}}}, \bibinfo{author}{\bibfnamefont{J.}~\bibnamefont{{Brinkmann}}}, \bibinfo{author}{\bibfnamefont{C.-H.} \bibnamefont{{Chuang}}}, \bibinfo{author}{\bibfnamefont{K.~S.} \bibnamefont{{Dawson}}}, \bibnamefont{et~al.}, \bibinfo{journal}{\mnras} \textbf{\bibinfo{volume}{499}}, \bibinfo{pages}{210} (\bibinfo{year}{2020}), \eprint{2007.08999}.

\bibitem[{\citenamefont{Ata et~al.}(2017)\citenamefont{Ata, Baumgarten, Bautista, Beutler, Bizyaev, Blanton, Blazek, Bolton, Brinkmann, Brownstein et~al.}}]{SDSSIVquasars}
\bibinfo{author}{\bibfnamefont{M.}~\bibnamefont{Ata}}, \bibinfo{author}{\bibfnamefont{F.}~\bibnamefont{Baumgarten}}, \bibinfo{author}{\bibfnamefont{J.}~\bibnamefont{Bautista}}, \bibinfo{author}{\bibfnamefont{F.}~\bibnamefont{Beutler}}, \bibinfo{author}{\bibfnamefont{D.}~\bibnamefont{Bizyaev}}, \bibinfo{author}{\bibfnamefont{M.~R.} \bibnamefont{Blanton}}, \bibinfo{author}{\bibfnamefont{J.~A.} \bibnamefont{Blazek}}, \bibinfo{author}{\bibfnamefont{A.~S.} \bibnamefont{Bolton}}, \bibinfo{author}{\bibfnamefont{J.}~\bibnamefont{Brinkmann}}, \bibinfo{author}{\bibfnamefont{J.~R.} \bibnamefont{Brownstein}}, \bibnamefont{et~al.}, \bibinfo{journal}{Monthly Notices of the Royal Astronomical Society} \textbf{\bibinfo{volume}{473}}, \bibinfo{pages}{4773–4794} (\bibinfo{year}{2017}), ISSN \bibinfo{issn}{1365-2966}, \urlprefix\url{http://dx.doi.org/10.1093/mnras/stx2630}.

\bibitem[{\citenamefont{Alam et~al.}(2017)\citenamefont{Alam, Ata, Bailey, Beutler, Bizyaev, Blazek, Bolton, Brownstein, Burden, Chuang et~al.}}]{BOSS}
\bibinfo{author}{\bibfnamefont{S.}~\bibnamefont{Alam}}, \bibinfo{author}{\bibfnamefont{M.}~\bibnamefont{Ata}}, \bibinfo{author}{\bibfnamefont{S.}~\bibnamefont{Bailey}}, \bibinfo{author}{\bibfnamefont{F.}~\bibnamefont{Beutler}}, \bibinfo{author}{\bibfnamefont{D.}~\bibnamefont{Bizyaev}}, \bibinfo{author}{\bibfnamefont{J.~A.} \bibnamefont{Blazek}}, \bibinfo{author}{\bibfnamefont{A.~S.} \bibnamefont{Bolton}}, \bibinfo{author}{\bibfnamefont{J.~R.} \bibnamefont{Brownstein}}, \bibinfo{author}{\bibfnamefont{A.}~\bibnamefont{Burden}}, \bibinfo{author}{\bibfnamefont{C.-H.} \bibnamefont{Chuang}}, \bibnamefont{et~al.}, \bibinfo{journal}{Monthly Notices of the Royal Astronomical Society} \textbf{\bibinfo{volume}{470}}, \bibinfo{pages}{2617–2652} (\bibinfo{year}{2017}), ISSN \bibinfo{issn}{1365-2966}, \urlprefix\url{http://dx.doi.org/10.1093/mnras/stx721}.

\bibitem[{\citenamefont{du~Mas~des Bourboux et~al.}(2017)\citenamefont{du~Mas~des Bourboux, Le~Goff, Blomqvist, Busca, Guy, Rich, Yèche, Bautista, Burtin, Dawson et~al.}}]{Laforestsquasarscross}
\bibinfo{author}{\bibfnamefont{H.}~\bibnamefont{du~Mas~des Bourboux}}, \bibinfo{author}{\bibfnamefont{J.-M.} \bibnamefont{Le~Goff}}, \bibinfo{author}{\bibfnamefont{M.}~\bibnamefont{Blomqvist}}, \bibinfo{author}{\bibfnamefont{N.~G.} \bibnamefont{Busca}}, \bibinfo{author}{\bibfnamefont{J.}~\bibnamefont{Guy}}, \bibinfo{author}{\bibfnamefont{J.}~\bibnamefont{Rich}}, \bibinfo{author}{\bibfnamefont{C.}~\bibnamefont{Yèche}}, \bibinfo{author}{\bibfnamefont{J.~E.} \bibnamefont{Bautista}}, \bibinfo{author}{\bibfnamefont{E.}~\bibnamefont{Burtin}}, \bibinfo{author}{\bibfnamefont{K.~S.} \bibnamefont{Dawson}}, \bibnamefont{et~al.}, \bibinfo{journal}{Astronomy and Astrophysics} \textbf{\bibinfo{volume}{608}}, \bibinfo{pages}{A130} (\bibinfo{year}{2017}), ISSN \bibinfo{issn}{1432-0746}, \urlprefix\url{http://dx.doi.org/10.1051/0004-6361/201731731}.

\bibitem[{\citenamefont{Abbott et~al.}(2018)\citenamefont{Abbott, Abdalla, Alarcon, Allam, Andrade-Oliveira, Annis, Avila, Banerji, Banik, Bechtol et~al.}}]{DES_Y1}
\bibinfo{author}{\bibfnamefont{T.~M.~C.} \bibnamefont{Abbott}}, \bibinfo{author}{\bibfnamefont{F.~B.} \bibnamefont{Abdalla}}, \bibinfo{author}{\bibfnamefont{A.}~\bibnamefont{Alarcon}}, \bibinfo{author}{\bibfnamefont{S.}~\bibnamefont{Allam}}, \bibinfo{author}{\bibfnamefont{F.}~\bibnamefont{Andrade-Oliveira}}, \bibinfo{author}{\bibfnamefont{J.}~\bibnamefont{Annis}}, \bibinfo{author}{\bibfnamefont{S.}~\bibnamefont{Avila}}, \bibinfo{author}{\bibfnamefont{M.}~\bibnamefont{Banerji}}, \bibinfo{author}{\bibfnamefont{N.}~\bibnamefont{Banik}}, \bibinfo{author}{\bibfnamefont{K.}~\bibnamefont{Bechtol}}, \bibnamefont{et~al.}, \bibinfo{journal}{Monthly Notices of the Royal Astronomical Society} \textbf{\bibinfo{volume}{483}}, \bibinfo{pages}{4866–4883} (\bibinfo{year}{2018}), ISSN \bibinfo{issn}{1365-2966}, \urlprefix\url{http://dx.doi.org/10.1093/mnras/sty3351}.

\bibitem[{\citenamefont{Kazin et~al.}(2014)\citenamefont{Kazin, Koda, Blake, Padmanabhan, Brough, Colless, Contreras, Couch, Croom, Croton et~al.}}]{WiggleZ}
\bibinfo{author}{\bibfnamefont{E.~A.} \bibnamefont{Kazin}}, \bibinfo{author}{\bibfnamefont{J.}~\bibnamefont{Koda}}, \bibinfo{author}{\bibfnamefont{C.}~\bibnamefont{Blake}}, \bibinfo{author}{\bibfnamefont{N.}~\bibnamefont{Padmanabhan}}, \bibinfo{author}{\bibfnamefont{S.}~\bibnamefont{Brough}}, \bibinfo{author}{\bibfnamefont{M.}~\bibnamefont{Colless}}, \bibinfo{author}{\bibfnamefont{C.}~\bibnamefont{Contreras}}, \bibinfo{author}{\bibfnamefont{W.}~\bibnamefont{Couch}}, \bibinfo{author}{\bibfnamefont{S.}~\bibnamefont{Croom}}, \bibinfo{author}{\bibfnamefont{D.~J.} \bibnamefont{Croton}}, \bibnamefont{et~al.}, \bibinfo{journal}{Monthly Notices of the Royal Astronomical Society} \textbf{\bibinfo{volume}{441}}, \bibinfo{pages}{3524–3542} (\bibinfo{year}{2014}), ISSN \bibinfo{issn}{0035-8711}, \urlprefix\url{http://dx.doi.org/10.1093/mnras/stu778}.

\bibitem[{\citenamefont{{Adame} et~al.}(2025)\citenamefont{{Adame}, {Aguilar}, {Ahlen}, {Alam}, {Alexander}, {Alvarez}, {Alves}, {Anand}, {Andrade}, {Armengaud} et~al.}}]{2024arXiv240403002D}
\bibinfo{author}{\bibfnamefont{A.~G.} \bibnamefont{{Adame}}}, \bibinfo{author}{\bibfnamefont{J.}~\bibnamefont{{Aguilar}}}, \bibinfo{author}{\bibfnamefont{S.}~\bibnamefont{{Ahlen}}}, \bibinfo{author}{\bibfnamefont{S.}~\bibnamefont{{Alam}}}, \bibinfo{author}{\bibfnamefont{D.~M.} \bibnamefont{{Alexander}}}, \bibinfo{author}{\bibfnamefont{M.}~\bibnamefont{{Alvarez}}}, \bibinfo{author}{\bibfnamefont{O.}~\bibnamefont{{Alves}}}, \bibinfo{author}{\bibfnamefont{A.}~\bibnamefont{{Anand}}}, \bibinfo{author}{\bibfnamefont{U.}~\bibnamefont{{Andrade}}}, \bibinfo{author}{\bibfnamefont{E.}~\bibnamefont{{Armengaud}}}, \bibnamefont{et~al.}, \bibinfo{journal}{\jcap} \textbf{\bibinfo{volume}{2025}}, \bibinfo{eid}{021} (\bibinfo{year}{2025}), \eprint{2404.03002}.

\bibitem[{\citenamefont{{Akaike}}(1974)}]{1974ITAC...19..716A}
\bibinfo{author}{\bibfnamefont{H.}~\bibnamefont{{Akaike}}}, \bibinfo{journal}{IEEE Transactions on Automatic Control} \textbf{\bibinfo{volume}{19}}, \bibinfo{pages}{716} (\bibinfo{year}{1974}).

\bibitem[{\citenamefont{Schwarz}(1978)}]{10.1214/aos/1176344136}
\bibinfo{author}{\bibfnamefont{G.}~\bibnamefont{Schwarz}}, \bibinfo{journal}{The Annals of Statistics} \textbf{\bibinfo{volume}{6}}, \bibinfo{pages}{461 } (\bibinfo{year}{1978}), \urlprefix\url{https://doi.org/10.1214/aos/1176344136}.

\bibitem[{\citenamefont{Burnham and Anderson}(2004)}]{doi:10.1177/0049124104268644}
\bibinfo{author}{\bibfnamefont{K.~P.} \bibnamefont{Burnham}} \bibnamefont{and} \bibinfo{author}{\bibfnamefont{D.~R.} \bibnamefont{Anderson}}, \bibinfo{journal}{Sociological Methods \& Research} \textbf{\bibinfo{volume}{33}}, \bibinfo{pages}{261} (\bibinfo{year}{2004}), \eprint{https://doi.org/10.1177/0049124104268644}, \urlprefix\url{https://doi.org/10.1177/0049124104268644}.

\bibitem[{\citenamefont{Bevington~et al.}(1969)}]{Bevington1969}
\bibinfo{author}{\bibfnamefont{.}~\bibnamefont{Bevington~et al.}}, \emph{\bibinfo{title}{Data Reduction and Error Analysis for the Physical Sciences}} (\bibinfo{year}{1969}).

\bibitem[{\citenamefont{{Liddle}}(2004)}]{2004MNRAS.351L..49L}
\bibinfo{author}{\bibfnamefont{A.~R.} \bibnamefont{{Liddle}}}, \bibinfo{journal}{\mnras} \textbf{\bibinfo{volume}{351}}, \bibinfo{pages}{L49} (\bibinfo{year}{2004}), \eprint{astro-ph/0401198}.

\bibitem[{\citenamefont{{Odintsov} et~al.}(2024)\citenamefont{{Odintsov}, {S{\'a}ez-Chill{\'o}n G{\'o}mez}, and {Sharov}}}]{2024arXiv241209409O}
\bibinfo{author}{\bibfnamefont{S.~D.} \bibnamefont{{Odintsov}}}, \bibinfo{author}{\bibfnamefont{D.}~\bibnamefont{{S{\'a}ez-Chill{\'o}n G{\'o}mez}}}, \bibnamefont{and} \bibinfo{author}{\bibfnamefont{G.~S.} \bibnamefont{{Sharov}}}, \bibinfo{journal}{arXiv e-prints} \bibinfo{eid}{arXiv:2412.09409} (\bibinfo{year}{2024}), \eprint{2412.09409}.

\bibitem[{\citenamefont{Kumar et~al.}(2023)\citenamefont{Kumar, Nunes, Pan, and Yadav}}]{Kumar:2023bqj}
\bibinfo{author}{\bibfnamefont{S.}~\bibnamefont{Kumar}}, \bibinfo{author}{\bibfnamefont{R.~C.} \bibnamefont{Nunes}}, \bibinfo{author}{\bibfnamefont{S.}~\bibnamefont{Pan}}, \bibnamefont{and} \bibinfo{author}{\bibfnamefont{P.}~\bibnamefont{Yadav}}, \bibinfo{journal}{Phys. Dark Univ.} \textbf{\bibinfo{volume}{42}}, \bibinfo{pages}{101281} (\bibinfo{year}{2023}), \eprint{2301.07897}.

\bibitem[{\citenamefont{Basilakos et~al.}(2013)\citenamefont{Basilakos, Nesseris, and Perivolaropoulos}}]{PhysRevD.87.123529}
\bibinfo{author}{\bibfnamefont{S.}~\bibnamefont{Basilakos}}, \bibinfo{author}{\bibfnamefont{S.}~\bibnamefont{Nesseris}}, \bibnamefont{and} \bibinfo{author}{\bibfnamefont{L.}~\bibnamefont{Perivolaropoulos}}, \bibinfo{journal}{Phys. Rev. D} \textbf{\bibinfo{volume}{87}}, \bibinfo{pages}{123529} (\bibinfo{year}{2013}), \urlprefix\url{https://link.aps.org/doi/10.1103/PhysRevD.87.123529}.

\bibitem[{\citenamefont{Perivolaropoulos and Skara}(2023)}]{10.1093/mnras/stad451}
\bibinfo{author}{\bibfnamefont{L.}~\bibnamefont{Perivolaropoulos}} \bibnamefont{and} \bibinfo{author}{\bibfnamefont{F.}~\bibnamefont{Skara}}, \bibinfo{journal}{Monthly Notices of the Royal Astronomical Society} \textbf{\bibinfo{volume}{520}}, \bibinfo{pages}{5110} (\bibinfo{year}{2023}), ISSN \bibinfo{issn}{0035-8711}, \eprint{https://academic.oup.com/mnras/article-pdf/520/4/5110/54034587/stad451.pdf}, \urlprefix\url{https://doi.org/10.1093/mnras/stad451}.

\bibitem[{\citenamefont{Planck~Collaboration}(2020)}]{planckAghanim:2018eyx}
\bibinfo{author}{\bibfnamefont{.}~\bibnamefont{Planck~Collaboration}}, \bibinfo{journal}{A\&A} \textbf{\bibinfo{volume}{641}}, \bibinfo{pages}{A6} (\bibinfo{year}{2020}), \urlprefix\url{https://doi.org/10.1051/0004-6361/201833910}.

\bibitem[{\citenamefont{{Planck Collaboration} et~al.}(2020{\natexlab{b}})\citenamefont{{Planck Collaboration}, {Akrami}, {Arroja}, {Ashdown}, {Aumont}, {Baccigalupi}, {Ballardini}, {Banday}, {Barreiro}, {Bartolo} et~al.}}]{planckinflation2020A&A...641A..10P}
\bibinfo{author}{\bibnamefont{{Planck Collaboration}}}, \bibinfo{author}{\bibfnamefont{Y.}~\bibnamefont{{Akrami}}}, \bibinfo{author}{\bibfnamefont{F.}~\bibnamefont{{Arroja}}}, \bibinfo{author}{\bibfnamefont{M.}~\bibnamefont{{Ashdown}}}, \bibinfo{author}{\bibfnamefont{J.}~\bibnamefont{{Aumont}}}, \bibinfo{author}{\bibfnamefont{C.}~\bibnamefont{{Baccigalupi}}}, \bibinfo{author}{\bibfnamefont{M.}~\bibnamefont{{Ballardini}}}, \bibinfo{author}{\bibfnamefont{A.~J.} \bibnamefont{{Banday}}}, \bibinfo{author}{\bibfnamefont{R.~B.} \bibnamefont{{Barreiro}}}, \bibinfo{author}{\bibfnamefont{N.}~\bibnamefont{{Bartolo}}}, \bibnamefont{et~al.}, \bibinfo{journal}{\aap} \textbf{\bibinfo{volume}{641}}, \bibinfo{eid}{A10} (\bibinfo{year}{2020}{\natexlab{b}}), \eprint{1807.06211}.

\end{thebibliography}
\end{document}